\documentclass[11pt,a4paper]{article}

\usepackage{jheppub}
\usepackage{epsfig,latexsym,amsfonts,amsmath,amsthm,amssymb,amsbsy,multirow,slashed,color,mathrsfs,wasysym,textcomp,subfigure,wrapfig,comment}
\usepackage[dvipsnames]{xcolor}
\usepackage{hyperref}
\usepackage[T1]{fontenc} 
\hypersetup{colorlinks,
	linkcolor = blue,
	anchorcolor = red,
	citecolor = blue,
	filecolor = red,
	urlcolor = blue,
	linktocpage}

\usepackage{benmacros}

\allowdisplaybreaks

\title{Mass-deformed M2 branes in Stenzel space} 

\author{\'Oscar J. C. Dias${}^{\,a}$,}
\author{Gavin S. Hartnett${}^{\,a}$,}
\author{Benjamin E. Niehoff${}^{\,b}$,}
\author{Jorge E. Santos${}^{\,c}$}

\affiliation{${}^{\,a}$ STAG Research Centre and the School of Mathematical Sciences \\ University of Southampton \\ Highfield SO17 1BJ, United Kingdom}
\affiliation{${}^{\,b}$ Institute for Theoretical Physics \\ KU Leuven \\ Celestijnenlaan 200D, B-3001 Leuven, Belgium}
\affiliation{${}^{\,c}$ Department of Applied Mathematics and Theoretical Physics \\ University of Cambridge \\ Wilberforce Road, Cambridge CB3 0WA, United Kingdom}
\emailAdd{ojcd1r13@soton.ac.uk}
\emailAdd{g.s.hartnett@soton.ac.co.uk}
\emailAdd{ben.niehoff@kuleuven.be}
\emailAdd{jss55@cam.ac.uk}

\abstract{We obtain finite-temperature M2 black branes in 11-dimensional supergravity, in a $G_4$-flux background whose self-dual part approaches a solution of Cveti\v{c}, Gibbons, L\"u, and Pope, based upon Stenzel's family of Ricci-flat K\"ahler deformed cones.  Our solutions are asymptotically $AdS_4$ times a 7-dimensional Stiefel manifold $V_{5,2}$, and the branes are ``smeared'' to retain $SO(5)$ symmetry in the internal space.  The solutions represent a mass deformation of the corresponding dual $CFT_3$, whose full description is at this time only partially-understood.  We investigate the possibility of a confinement/de-confinement phase transition analogous to the $AdS_5 \times S^5$ case, and a possible Gregory-Laflamme type instability which could lead to polarised brane solutions which break $SO(5)$.  We discuss possible consequences for AdS/CFT and the KKLT cosmological uplift mechanism.}

\begin{document}
\maketitle

\section{Introduction\label{sec:intro}}
\noindent\indent One of the many surprises of the study of general relativity in higher dimensions is the non-uniqueness of black hole solutions.  Unlike the familiar four-dimensional context, where the Kerr-Newman solution is the unique stationary, asymptotically-flat black hole, gravity in higher dimensions admits a remarkable diversity of black hole solutions.  Further remarkable is that some of these black hole solutions have been found analytically, such as black rings, black Saturns, and systems of concentric black rings.  There is in fact strong evidence that there exist an infinite number of (stationary) black hole phases, distinguished not only by their non-spherical topologies, but also by the presence of non-uniform ripples along their horizons (sometimes referred to as ``lumpy black holes'').  Moreover, the exotic black holes of this wondrous m\'enagerie are not merely isolated points in solution space, but instead are interconnected by a web of linear instabilities and topology-changing transitions.\footnote{A useful reference collecting many of the major results for black holes in higher dimensions is Ref.~\cite{2012bhhd.book}.}

Much of this work on novel black holes has been done in the context of pure Einstein-Hilbert gravity in higher dimensions, and it is plausible that the inclusion of matter will only add to its richness.  Particularly interesting choices of matter are those stemming from 10- or 11-dimensional supergravity, especially for geometries which admit a holographic interpretation.  In such contexts black holes correspond to thermal states in the dual field theory, and knowledge of the different black hole solutions can be leveraged to construct the phase diagram for the field theory at strong coupling. In \cite{Dias:2015pda, Dias:2016eto} this program was carried out for the case of global $AdS_5 \times S^5$ solutions of IIB supergravity, dual to $\mathcal{N}=4$ SYM on the 3-sphere. In addition to the well-known ``large'' black hole solution Schwarzschild-$AdS_5 \times S^5$, the authors of \cite{Dias:2015pda, Dias:2016eto} also construct ``lumpy'' black holes (describing perturbations of the Schw-$AdS_5 \times S^5$ solution along the $S^5$), and localised black holes on the $S^5$ with horizon topology $S^8$.  An analysis of the free energy of the different solutions reveals that in the canonical ensemble the novel black holes are subdominant saddles and therefore do not alter what is already known about the Hawking-Page confinement/de-confinement transition \cite{Hawking:1982dh}. However, in the microcanonical ensemble the novel solutions do dominate at low energies, whereas at large energies the large black hole phase is dominant. The phase transition is first order, and the holographic interpretation of the localised black hole solution is a phase in the CFT where the $SO(6)$ R-symmetry is spontaneously broken to $SO(5)$.\footnote{Similar lumpy and localized thermal phases do exist on (1+1)-dimensional SYM theory on $\mathbb{R}_t\times S^1$ dual to D1 branes with compactified worldvolume. In this case, localized thermal phases dominate both the microcanononical and canonical ensembles at low energy \cite{Dias:2017uyv}.}

In this paper we will investigate black hole solutions in 11-dimensional supergravity which are asymptotitically $AdS_4 \times V_{5,2}$, where $V_{5,2} \equiv SO(5)/SO(3)$ is a 7-dimensional Einstein space known as a Stiefel manifold.\footnote{In fact $V_{5,2}$ is Sasaki-Einstein, meaning that it is Einstein and a cone over it is K\"ahler.  It can also be written as a $U(1)$ bundle over the (K\"ahler) Grassmannian $SO(5) / (SO(3) \times SO(2))$.}  The black holes we shall construct are planar with respect to the $AdS_4$ factor and have flux with a non-zero Chern-Simons term, and thus they serve as interesting extensions of the study of black holes in higher dimensions. More importantly, we can use these black holes to map out the phase diagram for the dual field theory, and study both confinement and spontaneous symmetry breaking, just as was done for the asymptotically $AdS_5 \times S^5$ solutions. Unfortunately the dual field theory of asymptotically $AdS_4 \times V_{5,2}$ solutions is much less well-understood than $\mathcal{N} = 4$ SYM. The authors of Ref.~\cite{Martelli:2009ga} have proposed that the field theory dual to such solutions is an $\mathcal{N} = 2$ Chern-Simons quiver gauge theory with gauge group $U(N) \times U(N)$, and they have also studied the effect of a mass deformation.  Ref.'s \cite{Ceresole:1999zg, Jafferis:2009th} instead propose different field theory duals, but do not consider mass deformations.  While the mass deformation is our main interest here, we will have no comment on the validity of any of these proposals, and unfortunately we will have to proceed with a less-than-ideal understanding of the dual theory.

It is known that at zero temperature the field theory at hand exhibits confinement, and in this limit the gravitational dual is known. It is given by the Cveti\v{c}-Gibbons-L\"u-Pope (or CGLP) solution \cite{Cvetic:2000db}, which has $\mathcal{N} = 1$ supersymmetry in eleven dimensions. The CGLP solution is in many ways the 11-dimensional analog of 10-dimensional Klebanov-Strassler (KS) solution \cite{Klebanov:2000hb} of IIB supergravity. The KS solution geometrizes the duality cascade of the dual field theory, wherein the rank of the gauge group is successively reduced under RG flow. A similar phenomenon is thought to occur for the field theory dual of the CGLP solution, although in this case the cascade is less well-understood than in Klebanov-Strassler, particularly in the deep IR \cite{Martelli:2009ga}. 

By constructing black hole solutions with CGLP asymptotics, we hope to learn about the confinement/de-confinement transition, as well as any spontaneous symmetry breaking. In analogy with the $AdS_5 \times S^5$ case we expect different types of black hole solutions to exist. The simplest solution will preserve the same amount of symmetry in the compactified dimensions as the zero temperature CGLP solution, which is to say that spatial slices of the horizon will be topologically $\mathbb{R}^2 \times V_{5,2}$. This will correspond to the high temperature deconfined phase of the dual CFT.  Another possible class of solutions are black holes with different horizon topology, for example localised black holes with topology $\mathbb{R}^2 \times S^7$ or black ringoids with $\mathbb{R}^2 \times S^4 \times S^3$.  And a third possibility are ``lumpy'' black holes in an intermediate phase, having $\mathbb{R}^2 \times V_{5,2}$ topology but with  symmetry-breaking ripples along the $V_{5,2}$ part of the horizon.  In this paper we will construct the high-temperature, symmetric black hole. General arguments suggest the existence of localised black holes, and our numerical results seem to hint towards a possible ringoid solution; however the construction of either of these is beyond the scope of the current paper. We hope to address these in future work.

It would be very interesting to construct black holes in the KS geometry, as the dual field theory is well-studied and corresponds to a supersymmetric gauge theory undergoing a duality cascade. In Ref.~\cite{Aharony:2007vg, Bena:2012ek} KS black holes were constructed which preserved the $SU(2) \times SU(2)$ symmetry of the compactified dimensions. The more challenging task of building localised or lumpy black holes with KS asymptotics is obstructed by the fact that the KS solution is non-conformal in the UV. Equivalently, the bulk solution is not asymptotically $AdS_5$---the warp function instead exhibits logarithmic running in the radial direction. From a practical point-of-view, the logarithmic behaviour makes the problem dramatically more difficult because the dominant numerical methods used in constructing non-linear solutions to Einstein's equation are tailored to functions that are smooth over their domain, and behave poorly otherwise \cite{Wiseman:2011by,Dias:2015nua}.\footnote{Although the logarithm is smooth on any open set of the real line, numerically implementing the KS asymptotics requires that we capture the behaviour at $r=\infty$ where $\log$ is not differentiable.} From a theoretical point of view, the lack of a conformal fixed point in the UV complicates a holographic interpretation. By contrast, the CGLP solution is conformal in the UV, the bulk is asymptotically $AdS_4$, and there are no logarithms, making the problem of constructing black holes more amenable to numerics.

In addition to extending the study of black holes in higher dimensions and investigating phase transitions in strongly coupled field theories, another motivation for constructing black holes with CGLP asymptotics is, perhaps surprisingly, the question of whether string theory is capable of describing inflation. It is a non-trivial task to find inflationary, de-Sitter-like solutions to string theory. There are a few proposals, the most popular of which is KKLT \cite{Kachru:2003aw}, wherein $AdS$ vacua of string theory are lifted into $dS$ vacua through the addition of anti-branes and corrections to the superpotential from various sources.\footnote{To be more precise, the key ingredient is that the added branes carry charge opposite to the flux background, which itself is able to carry charge because of the Chern-Simons couplings. In this paper we will consider the CGLP solution which only exists for one sign of the flux. The charge of any branes or black holes in this background will then be measured relative to the charge of this flux.} When the anti-branes are added to the flux background, they puff up into a polarized configuration that is metastable \cite{Kachru:2002gs}; however, this step in the KKLT proposal has been heavily scrutinised in recent years.  When constructing the supergravity solution corresponding to the backreaction of smeared anti-D3 branes added to the KS background, it has been observed that unfamiliar singularities arise whose interpretation is unclear \cite{Bena:2009xk}.\footnote{In the anti-brane literature a crucial distinction is the difference between localised anti-branes, which are point-like in the transverse dimensions, and smeared branes, which have been distributed uniformly along some or all of the transverse dimensions. We shall use this same terminology to describe black holes. Localised black holes will have spherical horizon topologies in the transverse dimensions and smeared black holes will preserve the same symmetry in the compactified dimensions as the original flux background. The high-temperature CGLP black hole with horizon spatial slice $\mathbb{R}^2 \times V_{5,2}$ is therefore a smeared black hole.} It was thought by some that the singularities indicated an inconsistency or pathology in the KKLT proposal, and thus began a research program aimed at investigating this further.

A comprehensive review of this anti-brane problem would be too lengthy and cumbersome to give here; we will instead provide a brief discussion of the relevance of CGLP black holes to this problem. It was found in many explicit cases that singularities arise when anti-branes (smeared or localised) are added to flux backgrounds, and this was proved for a large class of flux backgrounds in \cite{Blaback:2014tfa}.  Specific studies showing that both smeared and localised anti-M2 branes added to the CGLP background are singular are Ref.s~\cite{Bena:2010gs, Massai:2011vi, Cohen-Maldonado:2016cjh}. Since the singularities are legitimate and not artefacts of any approximation, the key question then concerns the interpretation of these singularities and whether they are physical or not. One way to assess this is to regulate the singularities and examine their behaviour as the regulator is removed. A simple way to regulate the solutions is to shield the singularities with finite-radius black hole horizons. For flux backgrounds that admit a holographic interpretation, the field theory dual of an anti-brane added to a flux background is a metastable vacuum with broken supersymmetry. In this case the black hole regulator corresponds to considering this metastable state at finite temperature. In certain flux backgrounds it has been shown that \textit{smeared} black hole solutions with negative charge do not exist, meaning that the corresponding smeared anti-brane singularities cannot be regulated in this way \cite{Bena:2012ek, Bena:2013hr}. In Ref.~\cite{Hartnett:2015oda} one of us demonstrated that for certain \textit{localised} anti-branes, the singularities could in fact be shielded by horizons, indicating that a key factor was whether the anti-branes were localised or smeared. While not conclusive, this result provided some reassurance that anti-branes can in fact be used to create de Sitter solutions in string theory, since in the KKLT scenario localised anti-branes are used.\footnote{Ref.~\cite{Blaback:2014tfa} claims to rule out both localised anti-branes and localised negatively charged black holes in a large class of flux-backgrounds. We remain unconvinced by their argument in the finite temperature, black hole case. Rather than debate the issue here, we will merely note that this lack of agreement makes an exact analysis of localised black holes in CGLP even more desirable.}

The analysis in Ref.~\cite{Hartnett:2015oda} makes a few simplifying approximations and considered a simple toy flux background, and it would be desirable to examine localised black holes exactly in flux backgrounds such as CGLP (or KS) that are more directly relevant to the KKLT uplifting mechanism. If negatively charged localised black holes exist in these backgrounds, then this would mean that localised anti-brane singularities can indeed be shielded behind a horizon. The dual metastable states in the field theory would exist at finite temperature, and knowledge of the bulk solution might help facilitate a better understanding of the field theory state at zero temperature.\footnote{One thing to be mindful of when taking the zero temperature limit of black p-branes is that the behaviour of the temperature as extremality is approached depends on the particular solution in question. In particular, for the simple black p-branes of \cite{Horowitz:1991cd} the behaviour depends strongly the dimension and for some branes the temperature diverges as the extremal limit is approached.} 

With these motivations in mind, in this paper we take the first step towards the difficult task of constructing localised black holes in CGLP and mapping out the complete phase diagram for the mass deformed dual gauge theory. In Sec.~\ref{sec:11DonStenzel} we review the geometry of the Stenzel manifold and consider the dimensional reduction of 11 dimensional supergravity on this space. Then in Sec.~\ref{sec:review} we review the two analytically known solutions on this background, the planar Schwarzschild black hole and the CGLP solution itself. In Sec.~\ref{sec:perturbation} we perform a linear treatment as a warm up to the more complicated task of constructing the full, non-linear solution, which is done in Sec.~\ref{sec:nonlinear}. Lastly, we end with a concluding discussion. 

\section{11D Supergravity on squashed Stenzel space \label{sec:11DonStenzel}}
\subsection{Action and equations of motion \label{subsec:action}}
\noindent\indent Throughout this paper we will work in 11 dimensional supergravity, whose action is
\begin{equation}
S_{11D} = \frac{1}{2 \kappa_{11}^2} \int \bigg( R \hodge 1 - \frac{1}{2} \, G_{(4)} \wedge \hodge G_{(4)} + \frac16 \, G_{(4)} \wedge G_{(4)} \wedge A_{(3)} \bigg)\,,
\label{11d sugra action}
\end{equation}
where $G_{(4)}=\dd A_{(3)}$ and $R$ is the Ricci scalar (and $\hodge 1 = \vol_{11}$ is the volume form). The equations of motion are
\begin{subequations}
\begin{align}
&R_{ab} - \frac{1}{2} R\,g_{ab} = \frac12 T_{ab}\,,
\label{eq:einstein}
\\
&\dd \hodge G_{(4)} = \frac{1}{2} G_{(4)} \wedge G_{(4)},
\label{eq:maxwell}
\end{align}
\label{eqs:einsteinflux}
\end{subequations}
with energy-momentum tensor
\begin{equation}
T_{ab} = \frac{1}{6}\left[{G_{(4)}}_{acde} {G_{(4)}}_b{}^{cde} - \frac{1}{8} g_{ab} {G_{(4)}}_{cdef} {G_{(4)}}^{cdef}\right]\,.
\end{equation}
We will find it convenient, however, to express the Ricci scalar in terms of the trace of the energy-momentum tensor and thus move it to the other side,
\begin{equation}
R_{ab}= \frac{1}{12}\left[{G_{(4)}}_{acde} {G_{(4)}}_b{}^{cde} - \frac{1}{12} g_{ab} {G_{(4)}}_{cdef} {G_{(4)}}^{cdef}\right]\,.
\label{eq:einsteinreversed}
\end{equation}

We will be interested in a class of brane solutions of 11 dimensional supergravity for which the line element takes the general form 
\begin{equation}
\label{eq:metric-ansatz}
\dd s^2 = g_{tt} \, \dd t^2 + g_{x x} \left( \dd x_1^2 + \dd x_2^2 \right) + \dd s_8^2 \, ,
\end{equation}
where $\dd s_8^2$ is an 8-dimensional manifold that is topologically a Stenzel space (i.e. a deformed cone over the Stiefel manifold $V_{5,2}$ as in the CGLP solution \cite{Cvetic:2000db}), but on which certain squashing modes of the $V_{5,2}$ have been turned on.  The Ricci-flat Stenzel metrics of CGLP also have squashing modes of the $V_{5,2}$ turned on, as functions of the radial coordinate; however, in our solutions these squashing functions will be modified due to the presence of a black hole horizon.  Thus one might call our 8-dimensional transverse space a ``squashed Stenzel space''.  In any case, the 11-dimensional solutions are asymptotic to $AdS_4 \times V_{5,2}$ (each factor with its standard Einstein metric) and will be therefore dual to a strongly coupled large-$N$ gauge theory in $2+1$ dimensions. In the following section, we first briefly review the Stenzel manifold and the associated Stiefel manifold $V_{5,2}$, before discussing in more detail the class of solutions we will consider.

\subsection{The $n=3$ Stenzel space and the Stiefel manifold $V_{5,2}$ \label{subsec:stenzel}}
\noindent\indent The $n$-th Stenzel space is a $(2n+2)$-dimensional manifold homeomorphic to the cotangent bundle $T^*S^{n+1}$ of the sphere $S^{n+1} \simeq SO(n+2)/SO(n+1)$.  It is a complex manifold whose complex structure comes from a holomorphic embedding of $T^*S^{n+1}$ in $\CC^{n+2}$ given by the quadric surface,
\begin{equation}
\label{eq:quadric}
z^i z^i = \epsilon^2, \qquad \{z^i\} \in \CC^{n+2}, \qquad \epsilon \in \RR, \qquad i \in \{1,\dotsc,n+2\},
\end{equation}
and thus informally one can think of it as a hyperboloid asymptotic to the cone $z^i z^i = 0$, but deformed by blowing up an $S^{n+1}$ ``bubble'' in the center (at $x^i x^i = \epsilon^2$, where $x^i = \Re(z^i)$).  This hyperboloid can then be endowed with a metric which is K\"ahler and Ricci-flat.\footnotemark{}

\footnotetext{However, it is \emph{not} the metric which comes from the above embedding.}

The base of the asymptotic cone is not spherical, but is instead given by a different compact space, the Stiefel manifold $V_{n+2,2}$, and thus we will refer it as a ``Stiefel cone''.  Our particular case of interest is the 8-dimensional Stenzel space for $n=3$, and thus the base of the cone is $V_{5,2}$, which is the 7-dimensional coset $SO(5)/SO(3)$%
\footnote{For generic $p$ and $q$, the Stiefel manifold $V_{p,q} \equiv SO(p)/SO(p-q)$ is precisely the configuration space of $q$ oriented, orthonormal vectors in $p$-dimensional Euclidean space.}
where the $SO(3)$ is embedded in $SO(5)$ as the lower $3 \times 3$ block of a $5 \times 5$ matrix of $SO(5)$.%
\footnote{This is in contrast to the squashed 7-sphere, which is also an $SO(5)/SO(3)$ coset, but the $SO(3)$ is embedded in $SO(5)$ in a different way \cite{Castellani:1983tb}.}
It is also useful to think of $V_{5,2}$ as an $S^3$ bundle over $S^4$ (given by the bundle of unit cotangent vectors in $T^*S^4$)%
\footnote{Again, we note that this is \emph{not} the Hopf bundle, which would instead give us $S^7$ as its total space.};
then the ``tip'' of the Stenzel space is described by the $S^3$ fiber shrinking smoothly to zero size, leaving the finite-size $S^4$ to form a topological 4-cycle.

One can imagine that fully coordinatizing a space like $V_{5,2}$ can be complicated.  Since in this work we will preserve the $SO(5)$ symmetry, it will be more useful to write the metric in terms of left-invariant 1-forms of $SO(5)$.  Following \cite{Cvetic:2000db}, we begin with the standard $SO(5)$ Maurer-Cartan 1-forms $L_{AB}$, which are antisymmetric in $AB$, and where $A, B \in \{1,\dotsc,5\}$.  These forms satisfy the $SO(5)$ Lie algebra:
\begin{equation}
\dd L_{AB}=L_{AC}\wedge L_{CB}\,.
\label{eq:maurer}
\end{equation}
To parametrize the coset $SO(5)/SO(3)$, we will break the indices $A,B$ into the indices 1, 2, and $i,j \in \{3,4,5\}$, and then re-name a few of the $SO(5)$ forms as follows:
\begin{equation}
\sigma_i\equiv L_{1i}\,,\quad \tilde{\sigma}_i\equiv L_{2i}\,,\quad\text{and}\quad \nu\equiv L_{12}\,.
\end{equation}
The remaining $L_{ij}$ will then generate the $SO(3)$ on the lower diagonal block. The Maurer-Cartan relations (\ref{eq:maurer}) can now be written as
\begin{align}
\label{eq:left-inv-forms}
& \dd \sigma_i = \nu\wedge \tilde{\sigma}_i+L_{ij}\wedge \sigma_j\,,\quad \dd \tilde{\sigma}_i = -\nu\wedge \sigma_i+L_{ij}\wedge \tilde{\sigma}_j\,,\quad \dd \nu=-\sigma_i\wedge \tilde{\sigma}_i\nonumber
\\
& \dd L_{ij} = L_{ik}\wedge L_{kj}-\sigma_i \wedge\sigma_j-\tilde{\sigma}_i \wedge\tilde{\sigma}_j\,.
\end{align}
Expressions for these left-invariant forms in terms of explicit coordinates may be found in \cite{Bergman:2001qi} (and with a minor typo corrected in \cite{Klebanov:2010qs}).

Using the above left-invariant 1-forms, we can then write a ``round'' metric (i.e., with maximal isometry $SO(5) \times SO(2)$) on $V_{5,2}$ as follows:
\begin{equation}
\label{eq:stiefel-einstein}
\dd s^2(V_{5,2}) = \frac{9}{16}\nu^2 + \frac{3}{8} \sum_{i=1}^3 (\sigma_i^2 + \tilde{\sigma}_i^2 ) \,.
\end{equation}
Any choices of numerical factors would have preserved the maximal isometry; the particular choices $9/16$ and $3/8$ are such that the metric is Einstein, and its metric cone
\begin{equation}
\label{eq:stiefelcone}
\dd s^2_{C} = \dd r^2 + r^2 \dd s^2(V_{5,2}) \,
\end{equation}
is Ricci-flat.%
\footnote{In fact the metric \eqref{eq:stiefel-einstein} is Sasaki-Einstein, which means in addition that its metric cone is K\"ahler.}
This cone is a higher-dimensional analogue of the 6-dimensional conifold of \cite{Candelas:1989js} (in fact, the Stiefel manifold $V_{4,2}$ is the same as $T^{1,1}$).  There is of course a conical singularity at $r=0$, which can be smoothed out by blowing it up into a sphere (in our case an $S^4$), yielding the $n=3$ Stenzel space.  This is called ``deformation of the cone'', and lower-dimensional analogues include the deformed conifold ($n=2$), and the Eguchi-Hanson instanton ($n=1$).

We will focus on the case $n=3$, i.e. the 8-dimensional Stenzel space.  Using the left-invariant 1-forms of \eqref{eq:left-inv-forms}, one can write a cohomogeneity-1 ansatz for the Stenzel metric as follows:
\begin{equation}
\label{eq:stenzel}
\dd s^2_{S} = c^2 \left(\frac{\dd \tau^2}{4}  + \nu^2 \right) +\sum_{i=1}^3 \left( a^2\sigma_i^2+b^2 \tilde \sigma_i^2 \right) \,,
\end{equation}
where here $a$, $b$, and $c$ are functions of the radial coordinate $\tau$ only.  This ansatz of course includes the Stiefel cone \eqref{eq:stiefelcone} as a special case, for which 
\begin{equation}
a^2 = b^2 = \frac{3}{8} r^2, \qquad c^2 = \frac{9}{16} r^2, \qquad \tau = \frac{8}{3} \ln r \,.
\end{equation}
The ansatz \eqref{eq:stenzel} generically has $SO(5)$ symmetry, except in the case that $a^2 = b^2$, when that symmetry is enhanced to $SO(5) \times SO(2)$ (the $SO(2)$ in this case is the right action that rotates $\sigma_i$ into $\tilde \sigma_i$, originating from the $1,2$ indices of the original $L_{AB}$).  The functions $a,b,c$ are determined by the conditions that the metric be K\"ahler and Ricci-flat, and asymptotic to the Stiefel cone \eqref{eq:stiefelcone}.  We will work with the same conventions as \cite{Klebanov:2010qs}, where the solutions are written:
\begin{align}
\label{eq:stenzelfunctions}
a^2 &= \frac{3^{3/4} \ell_{\epsilon}^2}{2}  (2+\cosh \tau)^{1/4} \cosh(\tau/2) \,,  \nonumber \\
b^2 &= \frac{3^{3/4} \ell_{\epsilon}^2}{2}  (2+\cosh \tau)^{1/4} \frac{\sinh^2(\tau/2)}{\cosh(\tau/2)} \,, \nonumber \\
c^2 &= \frac{3^{7/4} \ell_{\epsilon}^2}{2}  \frac{\cosh^3(\tau/2)}{(2 + \cosh \tau)^{3/4} } \,.
\end{align}

Here $\ell_{\epsilon}$ is a length scale, sometimes also written as $\ell_{\epsilon} = \epsilon^{3/4}$.  In the large-$\tau$ limit, one recovers the Stiefel cone via the coordinate transformation
\begin{equation}
r^2 = \frac{2^{3/4}}{3^{1/4}} \, \ell_{\epsilon}^2 \, e^{3\tau/4} \,.
\end{equation}
However, as $\tau \to 0$, the function $b \to 0$, causing the $\tilde \sigma_i^2$ part of the metric \eqref{eq:stenzel} to be squashed to zero size.  This is in fact the $S^3$ fiber of $S^3 \hookrightarrow V_{5,2} \rightarrow S^4$ pinching off, and one can show that this happens smoothly, much like at the origin of $\RR^4$ in spherical coordinates.  This leaves an $S^4$ homology cycle at $\tau = 0$, as expected for a deformed conifold geometry, whose size is controlled by the length parameter $\ell_{\epsilon}^2$.

Ref.~\cite{Cvetic:2000db} showed that on this Stenzel space there exists a cohomology 4-form $\alpha_{(4)}$, dual to the homology 4-cycle, which is harmonic, normalisable, and self-dual%
\footnote{The notion of ``self-dual'' vs. ``anti-self-dual'' of course depends on the chosen orientation.}
$\alpha_{(4)} = \hodge_8  \alpha_{(4)}$, given by
\begin{eqnarray}
\label{eq:beta3}
&& \alpha_{(4)} = \dd \, \beta_{(3)}, \qquad \beta_{(3)} = \varepsilon_{ijk} \left( \frac{p}{6} \tilde{\sigma}_i \wedge \tilde{\sigma}_j \wedge \tilde{\sigma}_k + \frac{q}{2} \sigma_i \wedge \sigma_j \wedge \tilde{\sigma}_k \right) \, ,
\nonumber\\
&& p = \frac{9}{8} \frac{(1+3\cosh(\tau))}{\cosh^3(\tau/2)} \,, \qquad q = \frac{9}{4} \text{sech}(\tau/2) \,.
\end{eqnarray}
Since it is self-dual, $\alpha_{(4)}$ provides a Chern-Simons source $\alpha_{(4)} \wedge \alpha_{(4)} = \norm{\alpha_{(4)}}^2$ for the right-hand side of \eqref{eq:maxwell}.

We will be interested in brane-flux solutions where the 8-dimensional metric $ds_8^2$ in \eqref{eq:metric-ansatz} takes the form of the Stenzel ansatz \eqref{eq:stenzel}; however the squashing functions $a,b,c$ will change in order to accommodate a black-hole horizon at some value of $\tau$.  Thus we will refer to this 8-dimensional manifold as a ``squashed Stenzel space''.  Most of the properties we have just described for Stenzel spaces will no longer hold:  At finite temperature, the squashed Stenzel spaces of this paper will not be K\"ahler, nor Ricci-flat; however, the squashing will remain $SO(5)$-invariant, and there will still exist a normalisable 4-form (but it will not be self-dual).

\subsection{Dimensional reduction \label{subsec:4D}}
\noindent\indent Since we are interested in a field-theory interpretation of our results, in this section we present the dimensional reduction of the above ans\"atze down to 4 dimensions (along the $t, x_1, x_2,$ and $\tau$ coordinates).  We will consider the following reduction ansatz:
\begin{subequations} \label{eqs:lift}
\begin{align}
&\dd s^2 =e^{-3 \alpha-3 \beta -\gamma }\,\dd s_4^2+L^2\left[\frac{9}{4}e^{2\gamma}\nu^2 + \frac{3}{2} \sum_{i=1}^3 \left( e^{2\alpha}\sigma_i^2 + e^{2\beta}\tilde{\sigma}_i^2 \right) \right] \,,
\\
& G_{(4)} = L^{-1} \, \dd C_{(3)} +L^3\varepsilon_{ijk} \, \dd \left(\frac{P}{6} \,\tilde{\sigma}_i\wedge \tilde{\sigma}_j\wedge\tilde{\sigma}_k+\frac{Q}{2} \,\sigma_i\wedge \sigma_j\wedge\tilde{\sigma}_k\right) \,,
\label{eq:liftG4}
\end{align}
\end{subequations}
where $C_{(3)}$ is a 3-form potential living entirely in $\cM_4$.  Although in the end our solution will depend only on the radial coordinate $\tau$, for the reduction we keep arbitrary dependence on the coordinates of the 4-dimensional space $\cM_4$.  The reduced action is
\begin{align}
S_{4D} =\frac{1}{2\kappa_4^2}\int_{\mathcal{M}_4} \Big[ \hodge_4 {}^{(4)}R -&\frac{1}{2} K_{IJ}(\phi) \, \dd \phi^I \wedge \hodge_4 \dd \phi^J- \frac{1}{2L^2} e^{9\alpha + 9 \beta + 3 \gamma} \dd C_{(3)} \wedge \hodge_4 \dd C_{(3)} \nonumber \\
&- \frac{16}{27 L^2} \dd \left(Q^2 - P Q \right) \wedge C_{(3)} - \hodge_4 U(\phi) \Big] \,,
\end{align}
The kinetic matrix is given by
\begin{equation}
K_{IJ}(\phi) = \left(
\begin{array}{ccccc}
15 & 9 & 3 &0 & 0
\\
9 & 15 & 3 & 0 & 0
\\
3 & 3& 3& 0 & 0
\\
0 & 0 & 0 & \frac{8}{27}e^{-6\beta} & 0
\\
0 & 0 & 0 & 0 & \frac{8}{9}e^{-4\alpha-2\beta}
\end{array}
\right) \,,
\end{equation}
and $\phi^{I}=\{\alpha,\beta,\gamma, P, Q\}$ with $I \in\{1,2,3,4,5\}$ are the scalar fields.  The gravitational constants are related by
\begin{equation}
\kappa _4^2\equiv \frac{\kappa_{11}^2}{(2L)^7 \vol(V_{5,2})} = \frac{\kappa _{11}^2}{27 \pi ^4 L^7}\,,
\end{equation}
where we have used that the volume of $V_{5,2}$ is given by $2^{-7} 3^3 \pi^4$ \cite{Bergman:2001qi}. The factor of $(2L)$ is the radius of the $V_{5,2}$ that appears in the 11-dimensional $AdS_4 \times V_{5,2}$ solution. The 4-dimensional potential is given by
\begin{multline}
L^2 \, U(\phi)=e^{-5 \alpha -5\beta-3 \gamma } \left[\frac{3}{2} e^{4 \gamma }+\frac{2}{3} \left(e^{2 \alpha }-e^{2 \beta }\right)^2-6 e^{2 \gamma } \left(e^{2 \alpha }+e^{2 \beta}\right)\right]
\\+ \frac{16}{81}e^{-9\alpha -9\beta -3\gamma} \left[e^{4 \alpha +2 \beta } (P-2 Q)^2 + 3 e^{6 \beta } Q^2 \right]\,.
\end{multline}
We have explicitly checked that this truncation is consistent. That is to say, any solution of the equations of motion derived from this action can be embedded in $11$ dimensional supergravity via the lift defined in Eqs.~(\ref{eqs:lift}).

This theory features a 3-form potential $C_{(3)}$ whose equation of motion is
\begin{equation}
\dd \left( e^{9\alpha + 9 \beta + 3\gamma} \hodge_4 \dd C_{(3)} - \frac{16}{27} (Q^2 - P Q) \right) = 0 \,.
\end{equation}
It is natural to dualise the 4-form field strength and introduce $\tilde{F}_0 = e^{9\alpha + 9\beta+3\gamma} \hodge_4 \dd C_{(3)}$, whose solution is
\begin{equation}
\tilde{F}_0 = F_0 + \frac{16}{27} \left(Q^2 - P Q \right) \,.
\end{equation}
with $F_0$ a constant.  We can then substitute $\dd C_{(3)} = - e^{-9\alpha - 9\beta - 3\gamma} \hodge_4 \tilde F_0$ back into the action to obtain\footnote{Up to boundary terms which we will ignore, as they will not be needed here.  Appropriate boundary terms can be fixed by, e.g. holographic renormalization.}
\begin{equation}
\label{eq:4d-action}
S_{4D} =\frac{1}{2\kappa_4^2}\int_{\mathcal{M}_4} \Big[ \hodge_4 {}^{(4)}R - \frac{1}{2} K_{IJ}(\phi) \, \dd \phi^I \wedge \hodge_4 \dd \phi^J - \hodge_4 V(\phi) \Big] \,.
\end{equation}
where the potential $V$ is now
\begin{equation}
V(\phi)= U(\phi) + \frac{9}{2L^2} e^{-9\alpha -9\beta -3\gamma} \left( \frac{F_0}{3} + \frac{16}{81} (Q^2-PQ)\right)^2 \,.
\end{equation}
If we choose $F_0 = 3$, then the potential approaches $-6/L^2$ for $\phi^I \to 0$, and thus $\cM_4$ is asymptotically locally $AdS_4$ with radius $L$.  Different choices of $F_0$ correspond to different values of the $AdS$ radius.

Since we are interested in a holographic interpretation of our solutions near the boundary where they approach $AdS_4 \times V_{5,2}$, it is useful to expand the action \eqref{eq:4d-action} to second order around the $AdS_4$ vacuum, which gives the following action for small fluctuations in the scalars:
\begin{equation}
S_{4D}^{(2)} = \frac{1}{2\kappa^2_4}\int_{AdS_4} \dd ^4 x\,\sqrt{-{}^{(4)}\bar{g}}\left[- \frac{1}{2}K_{IJ}(0) \, {}^{(4)}\nabla_\mu\phi^{I} {}^{(4)}\nabla^\mu \phi^J - \frac{1}{2} M_{IJ} \phi^I \phi^J \right] \, ,
\end{equation}
where the mass matrix $M_{IJ}$ is given by
%
%
\begin{equation}
L^2 M_{IJ} = \bigg(\frac{L^2}{2}\bigg) \frac{\delta^2 V}{\delta \phi^I \delta \phi^J} \bigg\rvert_{0} =
\begin{pmatrix}
 \dfrac{610}{3} &  \dfrac{650}{3} & 66 & 0 & 0 \\[2.0ex]
 \dfrac{650}{3} & \dfrac{610}{3} &  66 & 0 & 0 \\[1.3ex]
 66 &  66 & 30 & 0 & 0 \\[0.5ex]
 0 & 0 & 0 & \dfrac{32}{81} & -\dfrac{208}{81} \\[2.0ex]
 0 & 0 & 0 & -\dfrac{208}{81} & \dfrac{512}{81}
\end{pmatrix}\,.
\end{equation}
We would furthermore like to diagonalise the scalars and find their masses. To accomplish this, we first choose a basis of scalars which diagonalises $K(0)$, and then rescale this basis so that the kinetic term is canonically normalised to $\frac{1}{2} \delta_{IJ}$.  We can then make an orthogonal transformation of the rescaled scalars which diagonalises $M$.  When the dust settles, the final transformation is
\begin{equation}
\tilde{\phi}^I = S^{I}{}_{J}\phi^J \,,
\end{equation}
where
\begin{equation}
S=
\begin{pmatrix}
\dfrac{9}{\sqrt{7}} & \dfrac{9}{\sqrt{7}} & \dfrac{3}{\sqrt{7}} & 0 & 0 \\[2.0ex]
0 & 0 & 0 & \dfrac{1}{3}\sqrt{\dfrac{2}{3}} & -\sqrt{\dfrac{2}{3}} \\[2.0ex]
-\sqrt{\dfrac{3}{7}} & -\sqrt{\dfrac{3}{7}} & 2 \sqrt{\dfrac{3}{7}} & 0 & 0 \\[1.3ex]
-\sqrt{3} & \sqrt{3} & 0 & 0 & 0 \\[0.5ex]
0 & 0 & 0 & \dfrac{\sqrt{2}}{3} & \dfrac{\sqrt{2}}{3}
\end{pmatrix}\,.
\end{equation}
The canonically-normalised quadratic action then becomes
\begin{equation}
S_{4D}^{(2)} = \frac{1}{2\kappa^2_4}\int \dd ^4 x\,\sqrt{-{}^{(4)}\bar{g}}\left[-\frac{1}{2} \delta_{IJ}{}^{(4)}\nabla_\mu\tilde{\phi}^{I} {}^{(4)}\nabla^\mu \tilde{\phi}^{J} - \frac{1}{2}\tilde{M}^2_{IJ} \tilde{\phi}^I \tilde{\phi}^J \right]\,,
\end{equation}
with the masses
\begin{equation}
L^2 \tilde{M}_{IJ}^2 = \diag \left( 18,10,4,- \frac{20}{9}, -\frac{14}{9} \right) \,.
\end{equation} 

Since these are just minimally coupled scalars, we can determine their conformal dimensions easily.  Expanding the $\cM_4$ metric in Fefferman-Graham coordinates  $x^{\mu} = (z_{\text{FG}}, x^a)$, $a=0,1,2$, we obtain
\begin{equation}
\label{eq:FGmetric}
\dd s^2_4 = \frac{L^2}{z_{\text{FG}}^2} \dd z_{\text{FG}}^2  + \sum_{n=0}^{\infty} g_{ab}{}^{(p_n)}(x) \, \dd x^a \dd x^b \, z^{p_n}_{\text{FG}} \,,
\end{equation}
where here $p_n$ is an increasing sequence of (rational) powers, and the coefficients $g_{ab}{}^{(p_n)}(x)$ are metric tensors of $\sig{2}{1}$ signature. The leading term is given by $p_0 = -2$ with the flat Minkowski metric:
\begin{equation}
g_{ab}^{(p_0)} \dd x^a \dd x^b = - (\dd x^0)^2 + (\dd x^1)^2 + (\dd x^2)^2.
\end{equation}
We have introduced a general expression $p_n$ for the powers because for general matter content the expansion will involve non-integer powers of $z_{\text{FG}}$.

The boundary expansion of a mass $M$ scalar is 
\begin{equation}
\phi(z,x) \sim z_{\text{FG}}^{\Delta_+} \left( a_0(x) + a_1(x) z_{\text{FG}} + ... \right) + z_{\text{FG}}^{\Delta_-} \left( b_0(x) + b_1(x) z_{\text{FG}}  + ... \right) \, ,
\end{equation}
where
\begin{equation}
\label{eq:deltas}
\Delta_{\pm} = \frac{3 \pm \sqrt{9 + 4M^2 L^2}}{2} \,.
\end{equation}
If $M^2L^2 > M_{BF}^2L^2 + 1$ (where $M_{BF}^2 L^2 = -9/4$ is the four-dimensional Breitenl\"ohner-Freedman mass), one has $\Delta_- < 1/2$.  In that case a field with the $\Delta_-$ branch absent, i.e. $b_i(x) = 0$, corresponds to a normalisable fall-off, whereas the solution with $b_i(x) \neq 0$ is non-normalisable. The holographic interpretation is that $a_0(x)$ is a VEV for a dimension $\Delta_+$ operator in the CFT, and $b_0(x)$ is a source for that operator. This interpretation holds for the scalars $\tilde{\phi}^I$ for $I=1,2,3$, which have conformal dimensions $\Delta = 6, 5, 4$, respectively.

However, the $I=4,5$ scalars have masses in the range
\begin{equation}
\label{eq:masswindow}
M_{BF}^2L^2 < M^2L^2 < M_{BF}^2L^2 + 1 \,,
\end{equation}
for which $1/2<\Delta_{-}\leq\Delta_{+} <5/2$, and there is therefore an ambiguity in the boundary conditions. In this case both branches of solutions are normalisable. Setting the $b_i(x) = 0$ is known as standard quantisation, and setting the $a_i(x) = 0$ corresponds to alternate quantisation.\footnote{One may also have $a_0(x) \propto b_0(x)$ which corresponds to a multi-trace boundary condition. We will not consider such boundary conditions here.} As both boundary conditions are possible, we must resolve the ambiguity by considering the problem at hand. In our case, the requirement that the bulk and boundary theories be supersymmetric picks out a preferred boundary condition.  Exactly which of the two choices supersymmetry requires is dependent on the particular Kaluza-Klein mode in question.

First we consider the $\tilde \phi^5$ scalar.  The spectrum of conformal dimensions for supergravity perturbations of $AdS_4 \times V_{5,2}$ was worked out in \cite{Ceresole:1999zg}. As noted by \cite{Martelli:2009ga}, of the two possible dimensions for $\tilde{\phi}^5$, $\Delta = 2/3$ or $7/3$, only the second shows up in Ref.~\cite{Ceresole:1999zg}'s table of dimensions. Therefore we choose the conformal dimension $\Delta = 7/3$. This corresponds to standard quantisation. The coefficient of the $z_{\text{FG}}^{2/3}$ in the expansion for $\tilde{\phi}^5$ then corresponds to a source in the dual CFT.

Next, consider the $\tilde{\phi}^4$ scalar. We claim that this scalar receives alternate quantisation, so that $\Delta = 4/3$ is the conformal dimension of the dual operator. To confirm this explicitly, we could presumably do the same sort of analysis as in \cite{Ceresole:1999zg}. Unfortunately this seems to require a greater understanding of the Kaluza-Klein spectrum than we possess. We can infer the proper quantisation by examining the form of $\tilde{\phi}^4$ on a known supersymmetric supergravity solution which is not sourced by an operator of this form. This will be done in Sec.~\ref{subsec:CGLP} for the CGLP solution.

Having understood the boundary conditions for the 4D scalars, and taking the 4-metric to be asymptotically $AdS_4$, we can then work out the boundary conditions for the 11-dimensional fields. This will be done below in Sec.~\ref{subsec:boundaryconditions}.  (In fact the 4D analysis here is quite necessary in order to understand the appropriate 11D boundary conditions, since the two sets of fields are related in a non-trivial way.)

\section{Review of known solutions\label{sec:review}}

Our objective in this work is to construct what we will call the ``CGLP black brane'' solution of 11-dimensional supergravity, but it will be helpful to review two closely-related solutions which are known analytically: the Schwarzschild black brane, and the supersymmetric CGLP solution.

\subsection{Schwarzschild black brane \label{subsec:Schw}}
\noindent\indent The simplest brane solution we will consider is the AdS-Schwarzschild black brane, a Freund-Rubin solution of the form $\mathcal{M}_4 \times V_{5,2}$, where $\mathcal{M}_4$ is the planar $AdS_4$ Schwarzschild black hole: 
\begin{subequations}
\label{eq:schw}
\begin{equation}
\dd s^2 = \frac{L^2}{z^2} \left[- f(z) \dd t^2 + \frac{\dd z^2}{f(z)} + (\dd w_1^2+\dd w_2^2)\right] + 4L^2 \dd s^2(V_{5,2}) \,,
\label{eq:schwmetric}
\end{equation}
\begin{equation}
f(z) = \left(1-\frac{z^3}{z_+^3}\right) \,,
\end{equation}
\begin{equation}
A_{(3)} = -\frac{L^3}{z^3} f(z) \,\dd t\wedge\dd w_1\wedge \dd w_2\,.
\label{eq:schwform}
\end{equation}
\end{subequations}
This solution corresponds to the near-horizon geometry of finite-temperature M2-branes at the singular tip of the Stiefel cone \eqref{eq:stiefelcone}. The constant $z_+$ parametrises the temperature of the black brane via
\begin{equation}
L\,T=\frac{3}{4\pi z_+}\,.
\end{equation}
Here $L$ is the $AdS_4$ length scale, which is related to the number of M2-branes given by
\begin{equation}
\label{eq:SchwarzM2number}
N = \frac{1}{(2\pi \ell_p)^6} \int_{V_{5,2}} \hodge G_{(4)} =  \frac{81 \pi^4 L^6}{(2\pi \ell_p)^6} \, ,
\end{equation}
where $\ell_p$ is the 11-dimensional Planck length, related to the gravitational coupling via $4\pi \kappa^2_{11} = (2\pi \ell_p)^9$. 

\subsection{The CGLP solution \label{subsec:CGLP}}

In \cite{Cvetic:2000db}, Cveti\v{c}, Gibbons, L{\"u}, and Pope presented a smooth, supersymmetric (and thus zero-temperature) solution to 11-dimensional supergravity corresponding to ``fractional'' M2-branes dissolved in flux, which generalizes the Klebanov-Strassler solution \cite{Klebanov:2000hb} to higher dimensions.  This solution can be written as a warped product of $(2+1)$-dimensional Minkowski space and the 8-dimensional Stenzel space with $n=3$, as was discussed in Sec.~\ref{subsec:stenzel}.  The metric and 3-form potential take the form:
\begin{subequations}
\label{eq:generalansatz}
\begin{equation}
\label{eq:generalansatzmetric}
\dd s^2 = H^{-2/3} \dd x_{\mu} \dd x^{\mu} + H^{1/3} \dd s^2_S \,,
\end{equation}
\begin{equation}
\label{eq:generalansatz3form}
A_{(3)} = -H^{-1} \dd t \wedge \dd w^1 \wedge \dd w^2 + m \, \beta_{(3)} \, ,
\end{equation}
\end{subequations}
where $\dd x_{\mu} \dd x^{\mu}$ is $(2+1)$-dimensional Minkowski space, $\dd s^2_S$ is the Stenzel metric given in \eqref{eq:stenzelfunctions}, and $\beta_{(3)}$ was defined in \eqref{eq:beta3}. The function $H$ is given by \cite{Klebanov:2010qs}
\begin{equation}
\label{eq:cglpfunctions}
H = \frac{2^{3/2} 3^{11/4} m^2}{\ell_{\epsilon}^{6}}  \int^{\infty}_{(2+\cosh \tau)^{1/4}}  \frac{\dd \, t}{(t^4 - 1)^{5/2}} \,.
\end{equation}
The magnetic part of the 3-form potential preserves the $SO(5)$ symmetry of the Stenzel metric, and its corresponding magnetic 4-form $\alpha_{(4)} = \dd \beta_{(3)}$ is the (unique) $L^2$ normalisable harmonic form on this space.  Thus the ansatz \eqref{eq:generalansatz} has the right ingredients to be asymptotically $AdS_4 \times V_{5,2}$.

Re-writing the metric ansatz \eqref{eq:generalansatzmetric} in a $4+7$ split (where the 7-dimensional part is given by the $V_{5,2}$ directions), one obtains the 4-dimensional metric of Sec.~\ref{subsec:4D}:
\begin{equation}
\label{eq:CGLP4D}
\dd s^2_4 = e^{3\alpha + 3\beta + \gamma} \left[ H^{-2/3} \dd x_{\mu} \dd x^{\mu} + \frac{H^{1/3} c^2}{4} \dd \tau^2 \right] \,, 
\end{equation}
To analyze the asymptotics, it will be useful to introduce a Fefferman-Graham coordinate $z_{\text{FG}}$ defined by 
\begin{equation}
\label{eq:CGLPFG}
z_{\text{FG}} = 2^{5/4} 3^{1/4} \left( \frac{L}{\ell_{\epsilon}} \right)^2 e^{-3\tau/4} \left(1 - \frac{81}{26} e^{-\tau} + \frac{214515}{22984} e^{-2\tau} + ... \right) \,,
\end{equation}
for which the $\dd s_4^2$ takes the Fefferman-Graham form of \eqref{eq:FGmetric} with $L = m^{1/3}$. 

Since the CGLP asymptotics $AdS_4 \times V_{5,2}$ is the same as those of the Schwarzschild black brane, this means that their holographic dual theories have the same UV fixed point; the magnetic flux $\alpha_{(4)}$ (present in the CGLP case) then corresponds to a relevant deformation.  To understand the effect of the deformation, it is instructive to examine the M2-brane charge. Since the Chern-Simons term $G_{(4)} \wedge G_{(4)}$ is non-zero, the notion of charge becomes a bit nuanced in this background.  We must consider two distinct notions of charge, the Maxwell charge and the Page charge.\footnote{In \cite{Marolf:2000cb} three separate notions of charge are discussed for general supergravity configurations, but for the current context in which there are no localised brane sources, there are only two inequivalent definitions.} These are
\begin{subequations}
\begin{align}
Q_{\text{Maxwell}} &= \int_{V_{5,2}} \hodge G_{(4)} \,, \label{MaxQ} \\
Q_{\text{Page}} &= \int_{V_{5,2}} \left( \hodge G_{(4)} - \frac{1}{2} A_{(3)} \wedge G_{(4)} \right) \,. \label{MaxP}
\end{align}
\end{subequations}
The Maxwell charge includes contributions from the Chern-Simons term, which allows the flux to source itself; but as a consequence, it is not a conserved charge (since the charge within a Gaussian surface can receive volume contributions from the Chern-Simons term).  The Page charge, by contrast, is the integral of a conserved current, and is the appropriate quantity to quantise~\cite{Marolf:2000cb}:
\begin{equation}
N = \frac{Q_{\text{Page}}}{(2\pi \ell_p)^6} = \frac{81 \pi^4 m^2}{(2\pi \ell_p)^6} \in \mathbb{Z} \,.
\end{equation}
Note that with the identification $L = m^{1/3}$, already discussed below \eqref{eq:CGLPFG}, this is the same expression as in \eqref{eq:SchwarzM2number}, and thus this is the $N$ of the large-$N$ limit which relates the dual gauge theory to these gravitational solutions.

Following \cite{Martelli:2009ga, Klebanov:2010qs} we may define a running, or effective number of M2-branes using the Maxwell charge via
\begin{equation}
N(\tau) = \frac{Q_{\text{Maxwell}}(\tau)}{(2\pi \ell_p)^6} = \frac{81 \pi^4 m^2}{(2\pi \ell_p)^6} \tanh^4(\tau/2) \, .
\end{equation}
This is not quantised. In the UV, it agrees with the number of M2-branes computed via the Page charge, $N(\infty) = N$, and decreases monotonically away from the boundary, vanishing at $\tau=0$, the tip of the Stenzel manifold. The geometric effect of this flux is to smooth out the singularity of the Stiefel cone. The holographic interpretation of this running of $N(\tau)$ is a Seiberg-like duality cascade where the rank of the gauge groups decreases under RG flow \cite{Martelli:2009ga}. The flow leads to a confined state in the IR, with confinement scale $\ell_{\epsilon}$ (which can be removed via a rescaling of the Minkowski coordinates \cite{Klebanov:2010qs}). Because there is no horizon, the solution may be considered at any temperature upon Wick rotation to a Euclidean geometry. Thus, this CGLP solution is to be interpreted as the confined phase of the holographic theory, i.e. it plays a role analogous to that of thermal $AdS$ in the confinement/de-confinement transition of $AdS_5\times S^5$.

From the dimensionally reduced perspective, the 4-dimensional metric is not an Einstein space, and the scalars $\phi^I$ do not correspond to an extremum of $V$, i.e. $\delta V/\delta \phi^I \big|_{\text{CGLP}} \neq 0$.  Yet we may still apply the perturbative analysis in Sec.~\ref{subsec:4D} since the solution is asymptotically $AdS_4 \times V_{5,2}$.  Of particular interest are the boundary expansions for the tachyonic scalars $\tilde{\phi}^4$ and $\tilde{\phi}^5$. These have the boundary expansions
\begin{equation}
\tilde{\phi}^4 = - \sqrt{3}\left( \alpha - \beta \right)  = - 2^{-2/3} 3^{1/6} \left( \frac{\ell_{\epsilon}}{L} \right)^{8/3} z^{4/3}_{FG} + \mathcal{O}(z_{\text{FG}}^{8/3}) \,.
\end{equation}
\begin{equation}
\tilde{\phi}^5 = \frac{\sqrt{2}}{3} \left( P + Q \right) =  2^{2/3} 3^{5/6} \left( \frac{\ell_{\epsilon}}{L} \right)^{4/3} z_{\text{FG}}^{2/3} + \mathcal{O}(z_{\text{FG}}^{2}) \,.
\end{equation}
The conformal dimension of $\tilde{\phi}^5$ was previously found to be $\Delta = 7/3$, so the fall-off for this scalar corresponds to a source in the dual field theory. In particular it corresponds to a mass deformation. The conformal dimension for $\tilde{\phi}^4$ is less clear. The possibilities are $\Delta = 4/3$ or $5/3$. Because the only source involved in the CGLP solution is the mass deformation, this boundary condition for $\tilde{\phi}^4$ should not involve a sourcing term. We therefore infer that $\Delta = 4/3$ as the $\Delta = 5/3$ branch is absent. It is known that in Kaluza-Klein theory for 11 dimensional supergravity, low mass scalars sometimes receive alternate quantisation, see for example Ref.~\cite{Duff:1986hr}.

\section{Perturbative Analysis \label{sec:perturbation}}
\noindent\indent The CGLP black brane we wish to construct in this paper is a finite-temperature version of the CGLP solution.  It preserves the same $SO(5)$ isometry as CGLP, and therefore corresponds to the backreaction of smeared, finite-temperature branes.  One can think of the branes as being smeared over the homology $S^4$, but of course what actually happens is that the tip of the Stenzel space is replaced by a horizon of $V_{5,2}$ topology.

Before embarking on the task of building this full non-linear solution, we will first consider a perturbation of the Schwarzschild black brane corresponding to a linearised mass deformation. This should agree with the general solution linearised in the mass deformation parameter and will serve as a good seed for our numerics in the non-linear problem.%
\footnote{Note that we can perturbatively add flux to the Schwarzschild black brane, but we cannot perturbatively add a horizon to the CGLP solution because black holes are not small perturbations. Something like a matched asymptotic expansion would have to be preformed instead, a task we will not pursue here.}
Because the flux in the Schwarzschild black brane \eqref{eq:schw} is purely electric, it is consistent to perturb only the 3-form potential $\delta A_{(3)}$, leaving the metric fixed to linear order. Therefore, we will consider a linear 3-form perturbation of the form
\begin{equation}
\label{eq:generalansatzform}
\delta A_{(3)} = L^3 \, \epsilon^{ijk} \left( \frac{P}{6} \tilde{\sigma}_i \wedge \tilde{\sigma}_j \wedge \tilde{\sigma}_k + \frac{Q}{2} \sigma_i \wedge \sigma_j \wedge \tilde{\sigma}_k \right) \,.
\end{equation}

It is instructive to consider this perturbation from the 4-dimensional perspective. In this case, it merely corresponds to a subset of the perturbative analysis performed in Sec.~\ref{subsec:4D}. There we perturbed around a Freund-Rubin solution of the form $\mathcal{M}_4 \times V_{5,2}$, with $\mathcal{M}_4$ an asymptotically $AdS_4$ manifold. The Schwarzschild black brane fits this form, and so the analysis done in that section applies here as well. In this case $\mathcal{M}_4$ is the $AdS_4$ Schwarzschild black hole considered in Sec.~\ref{subsec:Schw}, and $\alpha = \beta = \gamma = 0$. Therefore, the 11-dimensional linearised 4-form equation 
\begin{equation}
\dd \hodge \delta G_{(4)} - G_{(4)} \wedge \delta G_{(4)} = 0 \,,
\end{equation}
is equivalent to the equations of motion for the diagonalised scalars $\tilde{\phi}^I$ worked out in Sec.~\ref{subsec:4D}, and restricted to the case that $\alpha = \beta = \gamma = 0$. The relationship between the $P,Q$ and the $\tilde{\phi}^I$ is:
\begin{align}
\tilde{\phi}^2 = \frac{\sqrt{2}}{3\sqrt{3}} (P-3Q), \qquad \tilde{\phi}^5 = \frac{\sqrt{2}}{3}(P+Q) \,.
\end{align}

We expect to find a regular solution of these equations corresponding to a mass deformation in the boundary CFT. Recall that $\tilde{\phi}^2$ has $M^2 L^2 = 10$, so that only one choice of fall-off is normalisable. It is not hard to verify that there is no regular non-zero solution with normalisable boundary conditions, so $\tilde{\phi}^2 = 0$. Alternatively, $\tilde{\phi}^5$ has a mass of $M^2 L^2 = -14/9$, and a source for this scalar corresponds to a mass deformation. Therefore, we search for a solution for $\tilde{\phi}^5$ that is regular at the horizon and falls off like $z_{\text{FG}}^{2/3}$ in Fefferman-Graham coordinates. Such a solution may be found analytically. In terms of the original $P,Q$ variables and the coordinates used in the Schwarzschild solution of Sec.~\ref{subsec:Schw} it is:
\begin{subequations}
\begin{equation}
P(z)=\mu\,z^{2/3} \,q(z)\,,\quad \text{and}\,\quad Q(z)=\frac{\mu}{3}\,z^{2/3}\,q(z)
\end{equation}
with
\begin{equation}
q(z)=\frac{\cos \left(\frac{\pi }{18}\right) \Gamma \left(\frac{4}{9}\right) \Gamma \left(\frac{7}{9}\right)^2}{\pi } \, _2F_1\left(\frac{2}{9},\frac{2}{9};1;1-z^3\right)\,.
\end{equation}
\end{subequations}
Here $\mu$ parametrizes the mass deformation, and the numerical prefactors have been chosen to ensure that $q(0)=1$. We will see that our full nonlinear numerical results will reproduce this expression when $\mu$ is small.

\section{The CGLP black brane \label{sec:nonlinear}}
\noindent\indent We now turn to the main challenge, the non-linear construction of the (smeared) CGLP black brane.  For the numerical construction of these solutions, it will be useful to employ the DeTurck trick in order to render the equations elliptic, which then makes them amenable to numerical methods.  We shall give a brief overview of the DeTurck method first, and then proceed to the details of our numerical solutions.

\subsection{The DeTurck method}
\label{subsec:deturck}

The essential problem which the DeTurck trick solves is that the Einstein equation has a great deal of gauge redundancy coming from diffeomorphism invariance.  The DeTurck trick, first introduced in \cite{Headrick:2009pv} and recently reviewed in \cite{Dias:2015nua}, is a method of gauge-fixing by making these extra degrees of freedom dynamical.  One accomplishes this by adding a new term to the Einstein equation of the form $\frac12 (\Lie_\xi g)_{ab} = \nabla_{(a} \xi_{b)}$, where $\xi^a$ is a vector field defined by the formula
\begin{equation}
\xi^a = g^{c b} \Big( \Gamma^a_{c b} - \bar \Gamma^a_{c b} \Big), \quad \text{or equivalently}, \quad \xi_a = g^{c b} \Big( \bar{\nabla}_c g_{ba} - \frac12 \bar \nabla_a g_{b c} \Big).
\label{xidef}
\end{equation}
The meaning of the quantities in this formula are as follows:  First one should choose a \emph{fixed} ``reference metric'' $\bar g_{ab}$ on our spacetime manifold $\cM_{11}$, which is essentially arbitrary, except that it should have the same isometries and asymptotic structure as the solutions we seek.  The reference metric serves as an initial ``guess'', and so it pays to choose a reference which is reasonably close to the expected solution.  Having chosen $\bar g_{ab}$, one then computes the connection $\bar \nabla$ which is Levi-Civit\`a with respect to $\bar g_{ab}$.  That is, $\bar \nabla$ preserves $\bar g_{ab}$, but \emph{not} $g_{ab}$, which is the unknown solution we are after.  One then sets $\xi^a$, as shown in \eqref{xidef}, equal to the trace of the difference of the two connections $\nabla, \bar \nabla$ (and thus $\xi^a$ is properly a tensorial object).

In doing this, one can show that the Lie derivative $\nabla_{(a} \xi_{b)}$ contains the second-order derivatives of the gauge degrees of freedom, which were previously missing from Einstein's equation.  So, one can then modify the equation by the addition of this term, which yields the Einstein-DeTurck equation,
\begin{equation}
R_{ab}-\nabla_{(a} \xi_{b)}= \frac{1}{12}\left[{G_{(4)}}_{acde} {G_{(4)}}_b{}^{cde} - \frac{1}{12} g_{ab} {G_{(4)}}_{cdef} {G_{(4)}}^{cdef}\right]\,,
\label{eq:turck}
\end{equation}
which is now properly elliptic.  Of course, one may be concerned that the problem has been re-defined by the addition of the new term, and this concern is not unwarranted:  It will become necessary to check that, on the solutions, the quantity $\chi = (\xi^a\xi_a)^{1/2}$ converges to zero sufficiently fast (at a rate dictated by the numerical method used).  Then when $\chi = 0$, the solutions of the Einstein-DeTurck equation \eqref{eq:turck} will in fact be solutions of the Einstein equation \eqref{eq:einsteinreversed}, in a gauge where the traces of the connection coefficients are numerically equal to those of the chosen reference metric $\bar g_{ab}$ (thus providing exactly the right number of gauge-fixing conditions).

One drawback with the DeTurck method is that one cannot fix a gauge ahead of time, because the gauge degrees of freedom have been made dynamical.  That is to say, one cannot choose very many \emph{specific} facts about the coordinates used to describe the metric, except for coordinates along isometries which are effectively removed from the problem.  Thus for our present cohomogeneity-one problem, the (metric) coefficient of the radial coordinate must include an arbitrary function to be solved for, as in $q(y) \, \dd y^2$, in order to leave the gauge unspecified.  Similarly, a cohomogeneity-two problem must include an expression for a completely-generic 2-dimensional metric, and so on.  For problems of low cohomogeneity, this drawback is far outweighed by the benefit attained in making the Einstein equation elliptic, but we know of no problems greater than cohomogeneity-three to which this method has been successfully applied with the computing resources available in the present day.

In the context of the present calculation with $SO(5)$ symmetry, properly reducing the equations to cohomogeneity-one is actually a non-trivial challenge.  This challenge comes from the proliferation of left-invariant one-forms $\nu, \sigma^i, \tilde \sigma^i$ in the metric ansatz for Stenzel space \eqref{eq:stenzel}.  While it is possible to write these out in coordinates, the expressions are quite long and calculations in a coordinate basis quickly become unwieldy.  But there is no good reason to use coordinates; it is far more efficient to use the Lie algebra \eqref{eq:left-inv-forms} directly, which ought to be sufficient for calculating all of the curvatures.  The authors, however, are not aware of any instance in the literature where the DeTurck method has been applied in a non-coordinate basis.  Therefore, we have developed in detail a method for doing so, which we give in Appendix~\ref{app:noncoord}.

\subsection{Specialized ansatz for numerics}
\label{subsec:numericsansatz}
\noindent\indent A convenient ansatz for the numerical construction of the CGLP black brane is
\begin{subequations}
\label{eq:ansatzfinal}
\begin{multline}
\dd s^2=L^2\left[\frac{-(1-y^9) y_+^2 \,q_1(y)\dd t^2+ y_+^2 \,q_3(y)(\dd w_1^2+\dd w_2^2)}{y^6}+\frac{9\,q_2(y)\dd y^2}{(1-y^9) y^2}\right] \\
+ L^2\left[\frac{9}{4}\,q_6(y)\nu^2 + \frac{3}{2} \sum_{i=1}^3 \left( q_4(y)\sigma_i^2 + \,q_5(y)\tilde{\sigma}_i^2 \right) \right] \,,
\label{eq:lineelementfinal}
\end{multline}
\begin{multline}
A_{(3)} = L^3\frac{(1-y^9) y_+^3 \,q_9(y)}{y^9}\,\dd t\wedge\dd w_1\wedge \dd w_2 \\ 
+ \frac{L^3\,y^2}{y_+^{2/3}}\varepsilon_{ijk}\left(\frac{1}{6}\,q_7(y)\,\tilde{\sigma}_i\wedge \tilde{\sigma}_j\wedge\tilde{\sigma}_k+\frac{1}{2}q_8(y)\,\sigma_i\wedge \sigma_j\wedge\tilde{\sigma}_k\right) \,,
\end{multline}
\end{subequations}
where $y=0$ is the location of the conformal boundary and the horizon is a null hypersurface with $y=1$.
Note the presence of $q_2(y)$ which is necessary to avoid prematurely fixing a gauge.  The full ansatz contains nine functions $q_i(y)$, with $i \in \{1,\dotsc,9\}$, to be determined by solving the combined system of \eqref{eq:maxwell} and \eqref{eq:turck}.  These nine functions must satisfy boundary conditions which will be detailed in Sec.~\ref{subsec:boundaryconditions}.  For the reference metric $\bar g_{ab}$ to be used in the DeTurck trick, we shall take the Schwarzschild black brane, which is described by $q_1,\dotsc,q_6 = 1$, $q_7 = q_8 = 0$, and $q_9 = 1$.

In fact, one may immediately integrate the $t, w_1, w_2$ component of the Maxwell equation \eqref{eq:maxwell} to obtain $q_9$ in terms of the other functions, thus reducing the number of unknowns to eight:
\begin{multline}
q_9(y) = \frac{y^9}{(1-y^9)} \int_{1}^{y}\Bigg\{\frac{9 \sqrt{q_1(u)} \sqrt{q_2(u)} q_3(u)}{u^{10} q_4(u){}^{3/2} q_5(u){}^{3/2} \sqrt{q_6(u)}}\\
-\frac{16 \sqrt{q_1(u)} \sqrt{q_2(u)} q_3(u) \left[q_7(u)-q_8(u)\right] q_8(u)}{9 u^6 q_4(u){}^{3/2} q_5(u){}^{3/2} \sqrt{q_6(u)}}\Bigg\}\, \dd u\,,
\end{multline}
where the integration constant has been chosen such that $q_9(0)=-1$ and $q_9(1)$ is finite. The remaining equations depend on $q_9$ only through the combination ${\partial_y [(1-y^9) q_9(y)/y^9]}$, and thus we can then eliminate $q_9$ \emph{algebraically} in terms of the above integrand, leaving us with eight coupled nonlinear ordinary differential equations for the remaining $q_1, \dotsc, q_8$.

\subsection{Boundary conditions \label{subsec:boundaryconditions}}
\noindent\indent We must choose boundary conditions at the horizon ($y=1$) and at asymptotic infinity ($y=0$).  At the horizon, we demand regularity of the line element \eqref{eq:lineelementfinal}, which requires $q_1(1)=q_2(1)$ and Robin boundary conditions for the remaining functions, which we will not reproduce here as they are rather long and uninteresting.  At infinity, one must choose the appropriate leading fall-offs, which in this case is a somewhat subtle matter.  While some of these fall-off conditions arise by straightforward matching to the the CGLP solution; others require comparison to the 4-dimensional reduction of Sec.~\ref{subsec:4D} in order to ensure that the mass deformation is turned on with no additional sources.

We will not derive the boundary conditions here, but merely present them.  Near $y=0$, we shall take all of our functions $q_i$ to be power series:
\begin{equation}
q_i(y)=\sum_{j=0}^{\infty}a_i^{(j)}y^j\,.
\end{equation}
We do not write any logarithmic terms, since we expect the asymptotic conformal symmetry to be preserved\footnote{See \cite{Taylor:2001pp} for a discussion of related systems where this is not the case.}. Most of the constants $a_i^{(j)}$ are functions of a few others that are free. The latter are completely fixed once a solution is found, \emph{i.e.} they will parametrise the expectation values in our boundary theory. Up to $\cO(y^8)$, the boundary expansions of the $q_i$ are given by:
\begin{align}
&q_1(y)=1-\frac{224 y^4 \mu ^2}{116883 y_+^{4/3}}+\mathcal{O}(y^8)\,,
\nonumber \\
&q_2(y)=1+ \frac{2272 \mu ^2 y^4}{116883 y_+^{4/3}}+\mathcal{O}(y^8)\,,
\nonumber \\
&q_3(y)= 1-\frac{224 y^4 \mu ^2}{116883 y_+^{4/3}}+\mathcal{O}(y^8)\,,
\nonumber \\
&q_4(y)=1+\theta _1 y^4+\beta  y^5+\mathcal{O}(y^8)\,,
\nonumber \\
&q_5(y)=1-\left(\frac{64 \mu ^2}{3159 y_+^{4/3}}+\theta _1\right) y^4 -\beta  y^5+\mathcal{O}(y^8)\,,
\\
&q_6(y)= 1-\frac{32 y^4 \mu ^2}{3159 y_+^{4/3}}+\mathcal{O}(y^8)\,,
\nonumber \\
&q_7(y)=\mu-\left(\frac{1552 \mu ^3}{350649 y_+^{4/3}}+\frac{7 \theta _1 \mu }{6}\right) y^4 + \theta _2 y^5+\mathcal{O}(y^8)\,,
\nonumber \\
&q_8(y)=\frac{\mu }{3}+\left(\frac{272 \mu ^3}{350649 y_+^{4/3}}-\frac{\mu  \theta _1}{6}\right) y^4+ \left(\frac{8 \beta  \mu }{39}+\frac{\theta _2}{3}\right) y^5+\mathcal{O}(y^8)\,, \nonumber
\end{align}
where $\theta_1$, $\theta_2$ and $\beta$ are functions of $y_+$ and $\mu$, and should be extracted once a solution is found. More constants, which we will detail below, appear at higher order in $y$ and they are all related to expectations values of the several scalars in the theory or the expectation value of the energy.

In the dimensional reduction of Sec.~\ref{subsec:4D} we found two scalars with mass-squared negative enough to support either  standard or alternative boundary conditions.  The scalar $\tilde \phi^5$ corresponds to a mass deformation of the dual field theory, and sourcing this field translates into the leading behaviour for $q_7, q_8$. The second scalar, $\tilde{\phi}^4$, comes from the Stenzel part of the metric, and should not be sourced.  The expansion for this scalar is
\begin{equation}
\tilde{\phi}^4(y)\propto \log(q_4)-\log(q_5)\approx \left(\frac{64 \mu ^2}{3159 y_+^{4/3}}+2 \theta _1\right) y^4+2 \beta  y^5+\mathcal{O}(y^6)\,.
\end{equation}
To eliminate the source, one must set the coefficient of $z_{\text{FG}}^{5/3}$ in Fefferman-Graham coordinates to zero.  The map to Fefferman-Graham coordinates is not very illuminating, so we omit it; the result is that we must set $\beta = 0$.

In order to implement these boundary conditions, it is more convenient to change to a different basis of functions $\hat q_i(y)$ given by
\begin{equation}
\begin{split}
&q_1(y)=1-\frac{224 y^4 \mu ^2}{116883 y_+^{4/3}}+y^7 \hat{q}_1(y)\,,
\\
&q_2(y)=1+ \frac{2272 \mu ^2 y^4}{116883 y_+^{4/3}}+y^7\hat{q}_2(y)\,,
\\
&q_3(y)= 1-\frac{224 y^4 \mu ^2}{116883 y_+^{4/3}}+y^7\hat{q}_3(y)\,,
\\
&q_4(y)=1+y^4\hat{q}_4(y)\,,
\\
&q_5(y)=1+y^4\hat{q}_5(y)\,,
\\
&q_6(y)= 1-\frac{32 y^4 \mu ^2}{3159 y_+^{4/3}}+y^7 \hat{q}_6(y)\,,
\\
&q_7(y)=\mu+y^3\hat{q}_7(y)\,,
\\
&q_8(y)=\frac{\mu}{3}+y^3\hat{q}_8(y)\,.
\end{split}
\end{equation}
Then in terms of the hatted quantities, the boundary conditions at $y=0$ become very simple:
\begin{equation}
\begin{gathered}
\hat{q}_1=\hat{q}_2=\hat{q}_3=\hat{q}_6=\hat{q}_7=\hat{q}_8=0\,, \\
\quad \hat{q}_4(0)+\hat{q}_5(0)+\frac{64 \mu ^2}{3159 y_+^{4/3}}=0\,,\quad\text{and}\quad\hat{q}^\prime_4(0)=0\,.
\end{gathered}
\end{equation}
The last of these boundary conditions expresses that $\beta =0$, i.e.\ that the scalar $\tilde{\phi}^4$ has alternative quantisation.  Of course, one must also rewrite the boundary conditions at the horizon in terms of the hatted variables; for instance the regularity condition $q_1(1)=q_2(1)$ becomes
\begin{equation}
\hat{q}_1(1)=\frac{64\,\mu^2}{2997\, y_+^{4/3}}+\hat{q}_2(y)\,.
\end{equation}
It is these hatted quantities that we will solve for numerically.

Finally, we need to understand how to compare quantities we calculate in this background (dual to a thermal, de-confined state) to quantities calculated in the CGLP background (dual to a confined state).  Thus we must understand the relationship between our CGLP black brane parameters $(L, y_+, \mu)$ and the CGLP parameters $(L, \ell_{\epsilon})$.  The two solutions differ in the IR, but they should agree in the UV to leading order.  Both solutions are asymptotically $AdS_4$ with radius $L$. The temperature is an IR quantity and does not affect the UV. Both solutions have magnetic 4-form flux turned on, and from the 4D perspective this corresponds to a source for the scalar $\tilde{\phi}^5$. By working in Fefferman-Graham coordinates for each solution, the source has the same normalisation in both solutions if we identify
\begin{equation}
\frac{4 \sqrt{2}}{9} L^{2/3} \mu = 2^{2/3} 3^{5/6} \left( \frac{\ell_{\epsilon}}{L} \right)^{4/3} \,.
\end{equation}

\subsection{Energy and a Smarr relation\label{subsec:energy}}
\noindent\indent Having sorted out the boundary conditions, we also need a definition of energy and a corresponding Smarr relation.  Since conformal symmetry remains intact in the UV, we expect the energy to be scale-invariant.  This has three important consequences: i) we can use scale invariance to fix $y_+=1$ and dial $\mu$ as we move in parameter space, since only the ratio $\hat{\mu}\equiv \mu/T^{2/3}$ is scale-invariant; ii) the scale invariance of the theory has important consequences for thermodynamics, such as allowing for a Smarr formula; iii) since the temperature has a thermodynamic conjugate variable (the entropy density) so will $\mu$. We will call this thermodynamic conjugate variable $\Theta$ (also a density).

From a physical perspective, $\mu$ is a parameter that we fix at infinity, and so we expect our energy density to be computed at fixed $\mu$. In the microcanonical ensemble, this means that the energy density $\rho$ should be a function only of the entropy density $s$ and $\mu$: $\rho = \rho(s,\mu)$. Scale invariance demands that this function be homogeneous. Since the mass dimension of the entropy density is $2$, the mass dimension of the energy density is $3$ and the mass dimension of $\mu$ is $2/3$, the density must behave in the following way under scale transformations:
\begin{equation}
\rho(\lambda^2 s,\lambda^{2/3}\mu)=\lambda^3\rho(s,\mu)\,.
\label{eq:scaling}
\end{equation}
We will see shortly that this scaling relation has remarkable consequences.  As we have just mentioned, the energy density is a function of $s$ and $\mu$ only, with conjugate variables $T$ and $\Theta$, and so we will have a first law of the form
\begin{equation}
\delta \rho = T \, \delta s + \Theta \, \delta \mu\,,
\label{eq:first}
\end{equation}
where $\delta$ denotes a variation along the moduli space of solutions; in our case, $\delta = \delta y_+\partial_{y_+}+\delta \mu\partial_{\mu}$. The scaling relation (\ref{eq:scaling}) and the first law (\ref{eq:first}) together imply a Smarr relation for the energy density:  If one differentiates \eqref{eq:scaling} with respect to $\lambda$, and then set $\lambda = 1$, one obtains the result
\begin{equation}
2 s \left(\frac{\partial \rho}{\partial s}\right)_{\mu}+\frac{2}{3}\mu\left(\frac{\partial \rho}{\partial \mu}\right)_{s}=3\rho(s,\mu) \quad \Rightarrow \quad  T s = \frac{3}{2}\rho-\frac{1}{3}\Theta \mu\,,
\end{equation}
where we have used the first law \eqref{eq:first} to obtain the final implication.

In a subsequent publication \cite{dias:2017a} we will show that one can associate a variation (in moduli space) of a conserved quantity to each Killing vector field $\hat{\xi}$ as follows. First, we can construct a closed $9-$form, which we coin $\omega_{\hat{\xi}}$, given by
\begin{equation}
\omega_{\hat{\xi}}=\omega_{\hat{\xi}}^{g}+\omega_{\hat{\xi}}^{A_{(3)}}\,,
\end{equation}
where
\begin{align*}
&h_{ab}\equiv\delta g_{ab}\,,\quad a_{(3)}\equiv\delta A_{(3)}\,,\quad \omega_{\hat{\xi}}^{g} =\frac{1}{2\kappa^2}\left[-\delta \dd \hat{\xi}-i_{\hat{\xi}}\hodge [(\nabla_b h^{b}_{\phantom{b}a}-\nabla_a h)e^a]\right]\,,
\\
&\omega_{\hat{\xi}}^{A_{(3)}} =\delta Q^{A_{(3)}}-i_{\hat{\xi}}\Theta^{A_{(3)}}\,,
\\
&\Theta^{A_{(3)}}_{\hat{\xi}}=\frac{1}{2\kappa^2}\left(\hodge G_{(4)}\wedge a_{(3)}-\frac{1}{3}G_{(4)}\wedge A_{(3)}\wedge a_{(3)}\right)\,,
\\
&Q^{A_{(3)}}_{\hat{\xi}}=-\frac{1}{2\kappa^2}\left(\hodge G_{(4)}-\frac{1}{3}G_{(4)}\wedge A_{(3)}\right)\wedge i_{\hat{\xi}}A_{(3)}\,.
\end{align*}
One can explicitly check, using the equations of motion and the fact that $\hat{\xi}$ is a Killing vector, that $\omega_{\hat{\xi}}$ is a closed $9-$form,
\begin{equation}
\dd \omega_{\hat{\xi}}=0\,.
\end{equation}

Our geometry is written in coordinates adapted to the three Killing directions along the worldvolume of the branes, whose Killing vectors are $\hat{\xi}=\partial_t$ and $\hat{\xi}=\partial_{w_i}$. We will denote hypersurfaces of constant $t$ and $w_i$ by $\Sigma_t$ and $\Sigma_{w_i}$. Since these hypersurfaces are 10-dimensional, we can integrate $\dd \omega_{\partial_t}$ and $\dd \omega_{\partial_{w_i}}$ along $\Sigma_t$ and $\Sigma_{w_i}$, respectively. For the hypersurfaces of constant $t$ we will assume that the coordinates $w_i$ are periodic with period $\Delta w_i$, and in the hypersurfaces of constant $w_i$ we will take $t\in[0,\Delta t]$. For either of these surfaces we have
\begin{equation}
0=\int_{\Sigma_{\hat{\xi}}} \dd \omega_{\hat{\xi}} = \int_{S^{y=1}_{\hat{\xi}}} \omega_{\hat{\xi}}-\int_{S^{y=0}_{\hat{\xi}}} \omega_{\hat{\xi}}, \quad \text{which implies} \quad \int_{S^{y=1}_{\hat{\xi}}} \omega_{\hat{\xi}}=\int_{S^{y=0}_{\hat{\xi}}}\omega_{\hat{\xi}}\,,
\label{eq:smarrdiff}
\end{equation}
where $S^{y=0,1}_{\hat{\xi}}$ is a constant-$y$ slice of $\Sigma_{\hat{\xi}}$ and we have used Stokes' theorem.

Using the boundary conditions of Sec.~\ref{subsec:boundaryconditions}, we can evaluate the $y=0$ slice of \eqref{eq:smarrdiff} for $\hat{\xi}=\partial_t$ to find
\begin{equation}
\frac{1}{\Delta w_1\Delta w_2}\int_{S^{y=0}_{\partial_t}}\omega_{\partial_t} = T \delta s\,.
\label{eq:bigone}
\end{equation}
Therefore it follows, by applying our first law \eqref{eq:first}, that at the $y=1$ slice of \eqref{eq:smarrdiff} one has
\begin{equation}
\frac{1}{\Delta w_1\Delta w_2}\int_{S^{y=1}_{\partial_t}}\omega_{\partial_t} = \dd \rho-\Theta \dd \mu\,.
\end{equation}
Finally we observe that the following difference yields $\delta(Ts)$:
\begin{equation}
\frac{1}{\Delta w_1\Delta w_2}\int_{S^{y=1}_{\partial_t}}\omega_{\partial_t}-\frac{1}{\Delta t}\int_{S^{y=1}_{\partial_{w_i}}}\omega_{\partial_{wi}} =\frac{1}{\Delta w_1\Delta w_2}\int_{S^{y=0}_{\partial_t}}\omega_{\partial_t}-\frac{1}{\Delta t}\int_{S^{y=0}_{\partial_{w_i}}}\omega_{\partial_{w_i}} = \delta (T s)\,,
\end{equation}
which implies, via the Smarr relation, that
\begin{equation}
\frac{1}{\Delta w_1\Delta w_2}\int_{S^{y=1}_{\partial_t}}\omega_{\partial_t}-\frac{1}{\Delta t}\int_{S^{y=1}_{\partial_{w_i}}}\omega_{\partial_{w_i}} =\delta \left(\frac{3}{2}\rho-\frac{1}{3}\Theta\,\mu\right)\,.
\label{eq:bigtwo}
\end{equation}

The left hand sides of \eqref{eq:bigone} and \eqref{eq:bigtwo} can be readily evaluated in terms of our asymptotic quantities. Furthermore, they provide two partial differential equations in $\mu$ and $y_+$ for two unknowns $\rho$ and $\Theta$, which can be solved up to a constant of integration $C_1$:
\begin{subequations}
\begin{align}
&\Theta = \frac{N^{3/2}}{4374 \pi ^6} \left[12 \pi  C_1 \mu ^{7/2}-5 y_+^{5/3} \hat{q}_7^{\prime\prime}(0)\right]\,,
\\
&\rho = \frac{N^{3/2}}{419904 \pi ^6} \left[256 \pi  C_1 \mu ^{9/2}-192 \mu  y_+^{5/3} \hat{q}_7''(0)-729 y_+^3 \left(3 \hat{q}_1''(0)-4\right)\right]\,.
\label{eq:ene}
\end{align}
\end{subequations}
The remaining integration constant can be fixed by requiring that the energy of the CGLP solution be zero.  This fixes $C_1 = 0$. However, let us imagine for a moment we did not have the CGLP solution. In this case we could still set $C_1$ to zero, by appealing to the fact that we expect the energy to be an analytic function of $\mu$.  This is best seen by the fact we can construct a perturbation theory around $\mu=0$ where $P$ and $Q$ only contain odd powers of $\mu$, and the remaining functions contain even powers of $\mu$ only.

Now that we have an energy density we can discuss some thermodynamic properties we would like to investigate. First, we point out that the energy density $\rho$ is not a scale-invariant quantity, so instead we will consider $\hat{\rho}=\rho/T^3$. Similarly, for the entropy density we should only consider $\hat{s}=s/T^2$. From these we can construct a dimensionless free energy density $\hat{f}\equiv \hat{\rho}-\hat{s}$. In terms of the scale-invariant quantities, the first law of thermodynamics takes a particularly simple form:
\begin{equation}
3\hat{f}(\hat{\mu})+\hat{s}(\hat{\mu})-\frac{2}{3}\hat{\mu}\hat{f}^\prime(\hat{\mu})=0\,.
\end{equation}
All the solutions we present in this work satisfy this form of the first law up to $10^{-6}\%$.

We will also find it informative to plot the specific heat density at constant $\mu$, which is defined as
\begin{equation}
c_\mu = T\left(\frac{\partial s}{\partial T}\right)_{\mu}\,.
\end{equation}
It is easy to see that $c_\mu$ is not a scale-invariant quantity, so we should consider $\hat{c}_\mu=c_\mu/T^2$ instead. In terms of our dimensionless quantitates, we find
\begin{equation}
\hat{c}_\mu = 2\hat{s}(\hat{\mu})-\frac{2}{3}\hat{\mu}\,\hat{s}^\prime(\hat{\mu})\,.
\end{equation}

Our final expression \eqref{eq:ene} for the energy does not agree with the one found in \cite{Cohen-Maldonado:2016cjh}. We have traced this discrepancy to the fact that \cite{Cohen-Maldonado:2016cjh} uses a subtraction method to deal with the AdS asymptotics, which is known to have ambiguities. Such ambiguities eventually lead to the birth of holographic renormalisation \cite{deHaro:2000xn}.

\subsection{Numerical results \label{subsec:results}}

\noindent\indent We are now finally ready to solve the system of equations for the CGLP black brane and discuss our results. In order to solve for the $8$ functions $\hat{q}_i$, with $i\in\{1,\ldots,8\}$, we discretise the system of coupled ordinary differential equations using a pseudo-spectral collocation method on a Chebyshev grid. The resulting nonlinear system of algebraic equations is then solved using a standard Newton-Raphson method algorithm. Once a solution is found, we compute its energy density via \eqref{eq:ene}. The entropy density can also be easily computed via the area of the horizon.

\begin{figure}[h]
\centering
\includegraphics[height=0.438\linewidth]{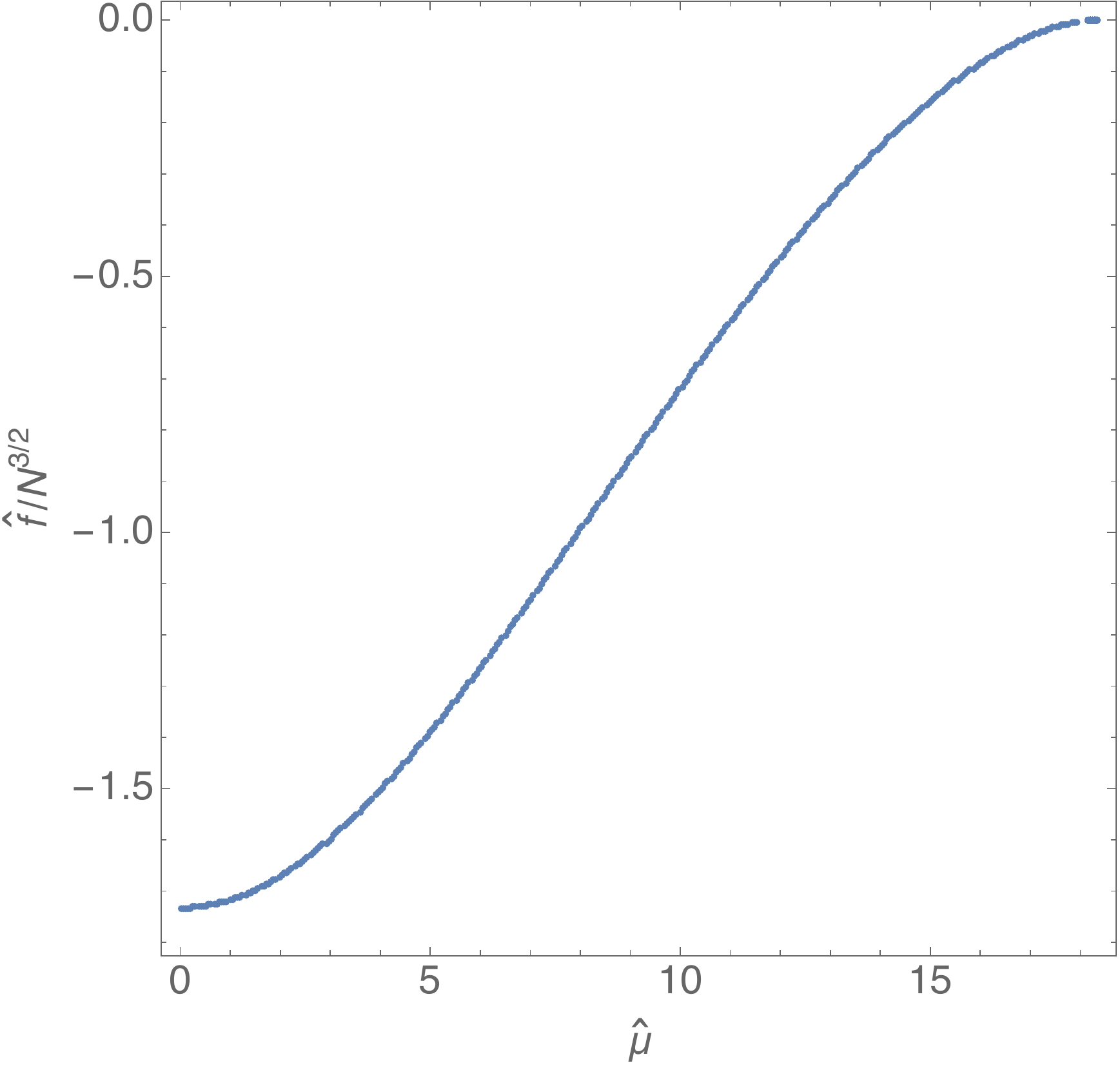}\hspace{1cm}
\includegraphics[height=0.438\linewidth]{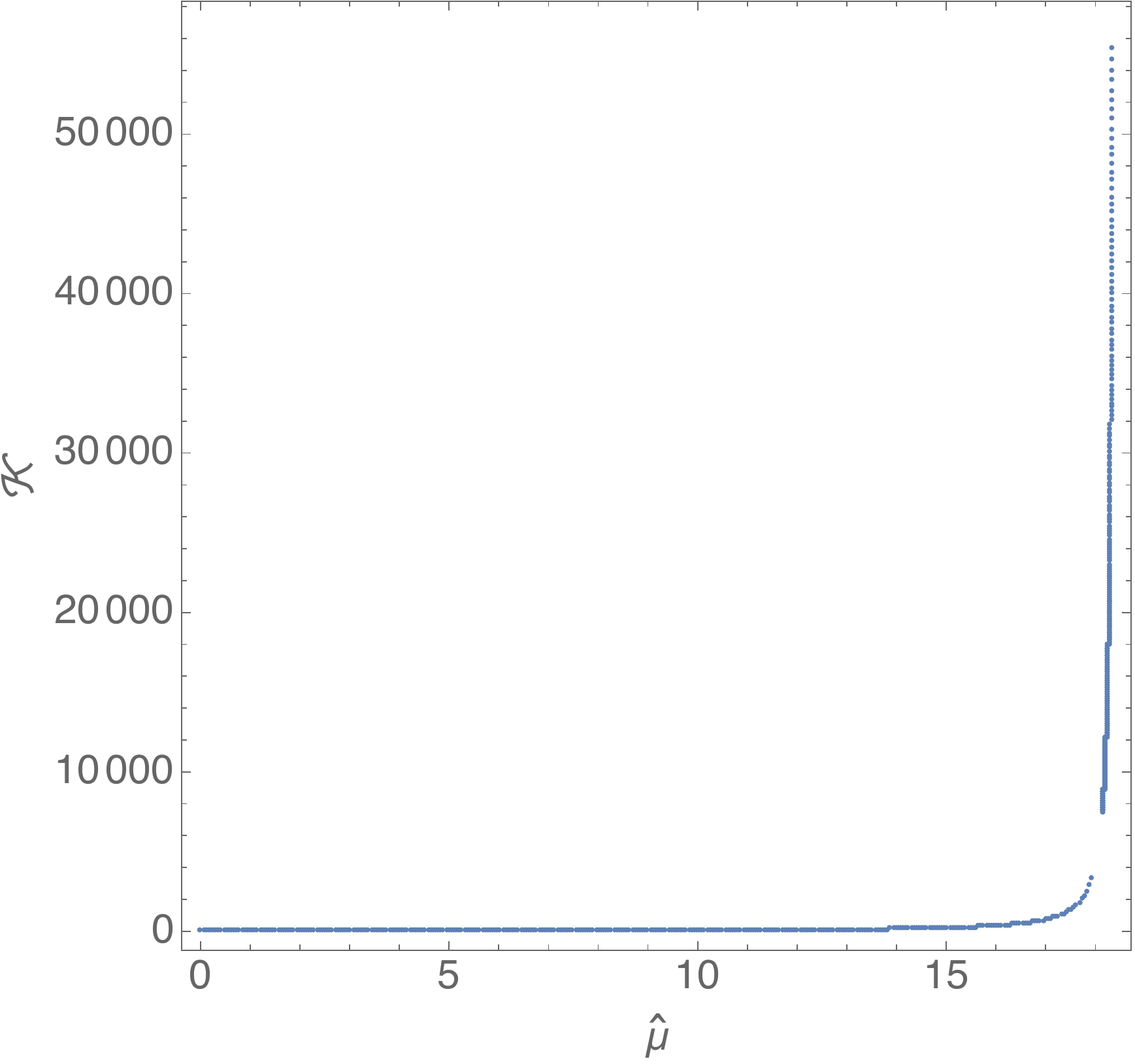}
\caption{{\bf Left Panel: }The dimensionless free energy density $\hat{f}$ as a function of $\hat{\mu}$: $\hat{f}$ reaches zero at $\hat{\mu}_c \approx 18.3906\pm 0.0046$ with what appears to be zero derivative. {\bf Right Panel: }The Kretschmann scalar $\mathcal{K}$ as a function of $\hat{\mu}$.}
\label{fig:freenergy}
\end{figure}

The most important quantity to plot is the dimensionless free energy $\hat{f}$. For small $\hat{\mu}$ (high temperature) $\hat{f}$ is negative, and vanishes for the CGLP solution, $\hat{f}_{\text{CGLP}} = 0$. Therefore, if $\hat{f}$ of the CGLP black brane crosses zero to become positive, it would signal a first order confinement/de-confinement phase transition in the field theory. On the gravity side, this would mean that the CGLP solution would have lower free energy than the black brane phase we constructed for sufficiently large $\hat{\mu}$ (small temperature). On the left panel of Fig.~\ref{fig:freenergy} we plot $\hat{f}$ as a function of $\hat{\mu}$. 

While we do not observe $\hat{f}$ crossing zero, we do find evidence for a phase transition occurring at
\begin{equation}
\hat{\mu}_{\mathrm{crit}}=\left(\frac{\mu}{T^{2/3}}\right)_{\mathrm{crit}}=18.3906\pm 0.0046\,.
\end{equation}
Perhaps surprisingly, this transition seems to be second order, since $\hat{f}$ has zero derivative as we approach $\hat{\mu}_{\mathrm{crit}}$ from below. An additional surprise is that the solution appears to be singular at the critical point. This can be seen in several ways. On the right panel of Fig.~\ref{fig:freenergy} we plot the Kretschmann scalar, $\mathcal{K}\equiv L^4 R^{abcd}R_{abcd}$, as a function of $\hat{\mu}$. We can measure, via a simple third order polynomial extrapolation, the value of $\hat{\mu}$ at which $\mathcal{K}$ diverges\footnote{This is best done by extrapolating the behaviour of $\mathcal{K}^{-1}$ as a function of $\hat{\mu}$.}, and that coincides with our estimate for $\hat{\mu}_{\mathrm{crit}}$ to within $0.02\%$!

\begin{figure}[h]
\centering
\includegraphics[height=0.438\linewidth]{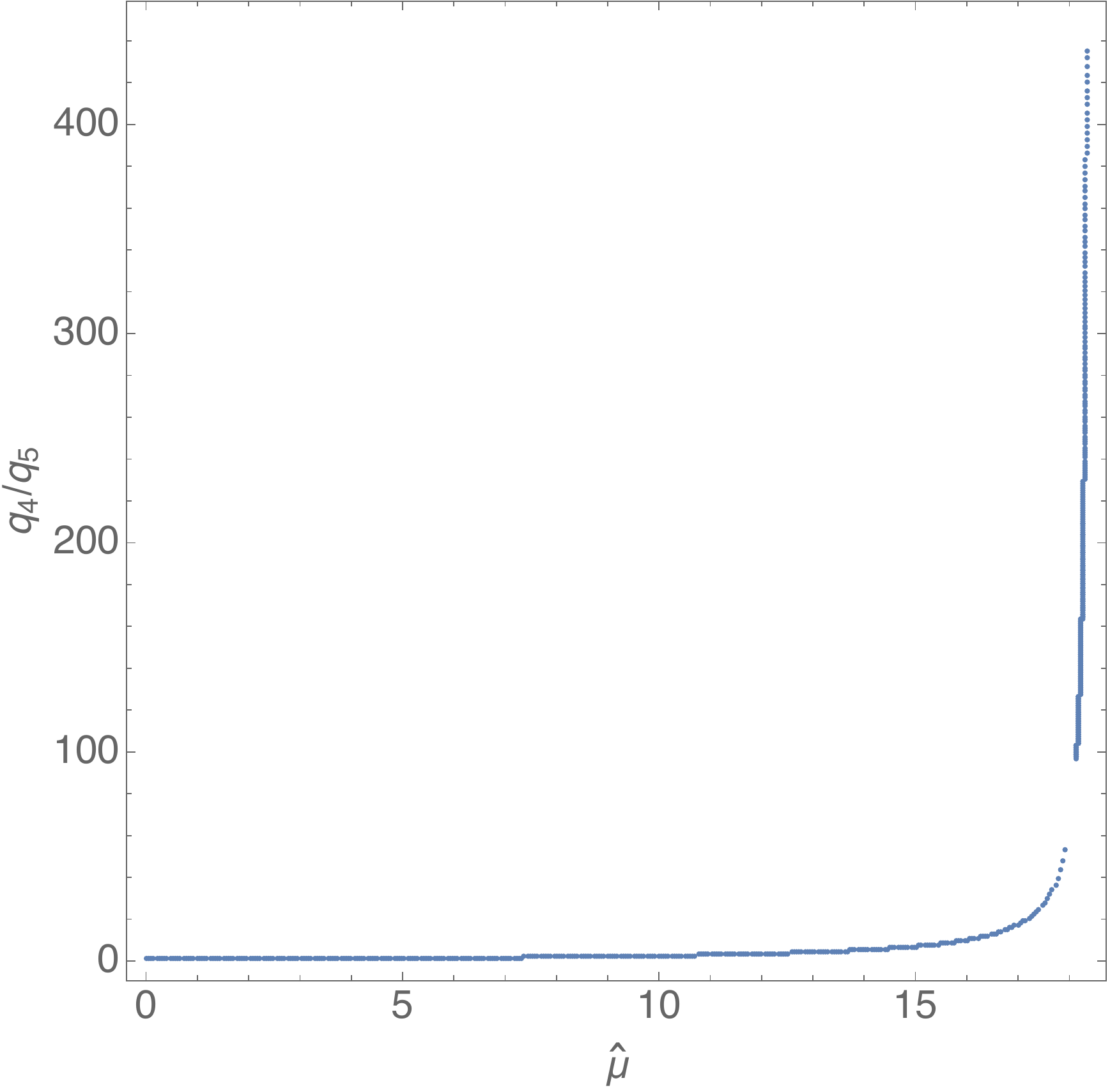}\hspace{1cm}
\includegraphics[height=0.438\linewidth]{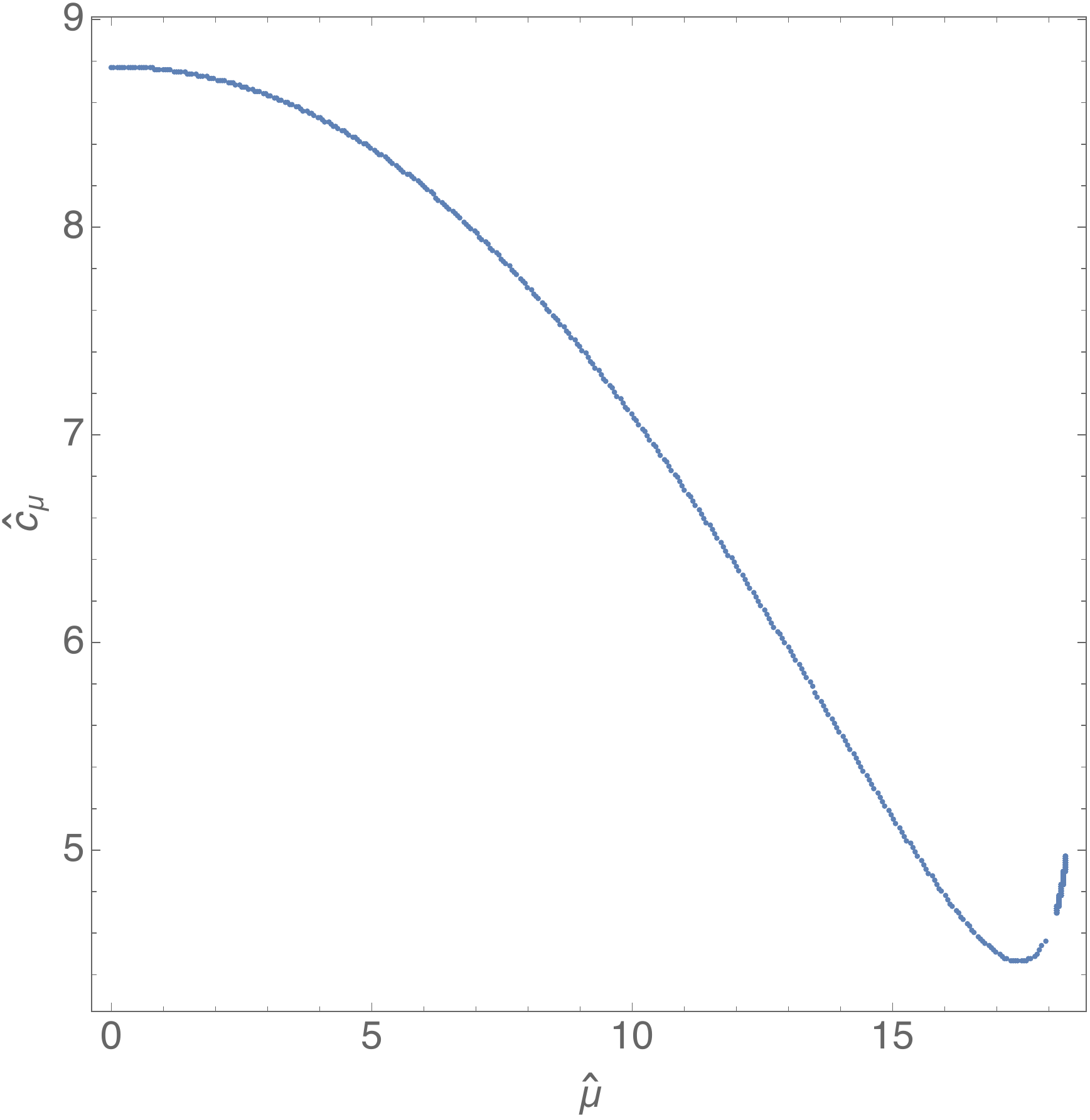}
\caption{{\bf Left Panel: }The ratio $(q_4/q_5)_{y=1}$ evaluated at the  as a function of $\hat{\mu}$. {\bf Right Panel: }The dimensionless specific heat $\hat{c}_\mu$ as a function of $\hat{\mu}$.}
\label{fig:ratio}
\end{figure}

In order to better understand the nature of the singularity we can study the ratio $q_4/q_5$. Evaluated on the CGLP solution, this ratio is $a/b$, which is 1 at the $AdS$ boundary and diverges like $a/b \sim  \tau^{-1}$ at the tip of the Stenzel metric. In the left panel of Fig.~\ref{fig:ratio} we plot the value of this ratio at the horizon. As $\hat{\mu}$ increases $q_4$ increases to about $7.563$ but seems to remain finite, while $q_5$ approaches zero very rapidly precisely at the location of the singularity, i.e. at $\hat{\mu}=\hat{\mu}_{\mathrm{crit}}$ (again, the extracted value via extrapolation agrees with $\hat{\mu}_{\mathrm{crit}}$ within $0.002\%$). By looking at our metric ansatz (\ref{eq:lineelementfinal}), it seems the $S^3$ described by the $\tilde{\sigma}_i$ wants to shrink at that point. We interpret this as evidence for a topological phase transition through a singular cone solution to a black brane with horizon topology $\mathbb{R}^2\times S^3\times S^4$. We plan to investigate this possibility in future work.

Next we comment on the specific heat at constant $\mu$, plotted in the right panel of Fig.~\ref{fig:ratio}. If this quantity were to change sign, it would indicate a Gregory-Laflamme type instability along the flat directions $w_1, w_2$ of the black brane we have constructed.  However, we find no evidence of this, as $\hat{c}_\mu$ does not change sign.

\begin{figure}[h]
\centering
\includegraphics[height=0.438\linewidth]{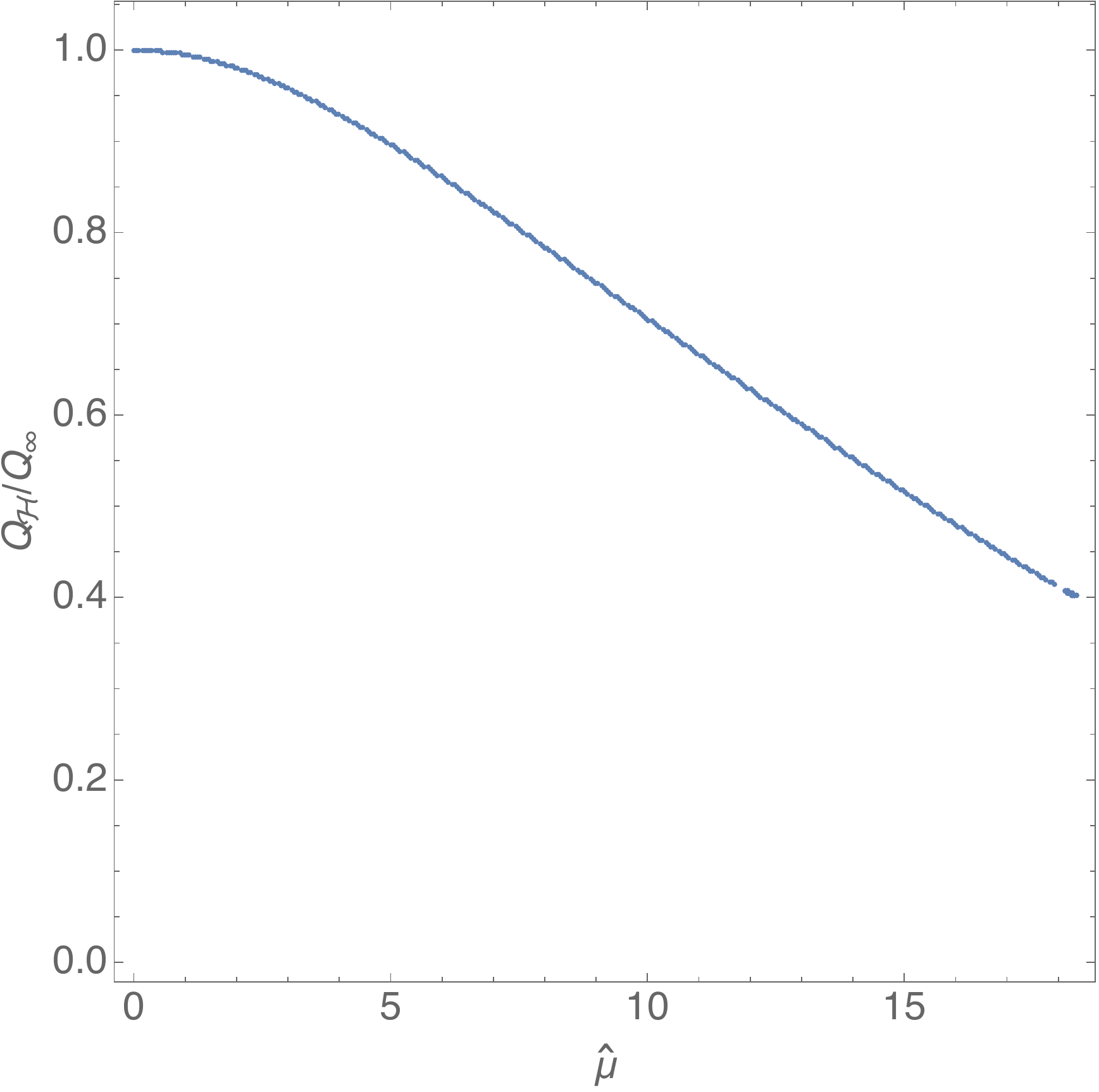}
\caption{Ratio of the Maxwell charge measured at the horizon to the Maxwell charge measured at infinity.}
\label{fig:chargeratio}
\end{figure}

Lastly, we will consider the M2-brane charge. Recall that in flux backgrounds where the Chern-Simons term is non-zero, $G_{(4)} \wedge G_{(4)} \neq 0$, we may consider both a Maxwell and a Page charge. The Page charge \eqref{MaxP} is conserved, and will take the same value in the solution considered here as in both the CGLP and Schwarzschild black brane solutions. In the current set-up the Page charge is just measuring the $AdS_4$ radius. The Maxwell charge \eqref{MaxQ} varies radially throughout the geometry, and agrees with the Page charge at infinity since the Chern-Simons term vanishes asymptotically. We therefore consider the ratio
\begin{equation}
\frac{Q_{\text{Maxwell}}(y)}{Q_{\text{Page}}} = \frac{Q_{\text{Maxwell}}(y)}{Q_{\text{Maxwell}}(0)} = \frac{N(y)}{N(0)} \,.
\end{equation}
Of particular interest is this ratio evaluated at the horizon, in which case it compares the M2-brane charge at the boundary to the M2-brane charge at the horizon. This quantity is of great relevance for understanding the fate of anti-brane singularities as discussed in the Introduction. If the charge at the horizon is the same sign as at infinity for all solutions, it means that the  singularities of smeared anti-branes added to the CGLP solution cannot be cloaked with a finite-temperature horizon. This is what has been found for a wide variety of smeared black holes in other flux backgrounds.\footnote{We should point out that an inability to blacken an anti-brane singularity does not necessarily rule it out as unphysical; for example in Ref.~\cite{Gubser:2000nd} it was noted that there are singular Coulomb branch solutions in $AdS_5/CFT_4$ which are physical and yet cannot support any finite temperature. Conversely, however, the ability to blacken a singularity \emph{does} support an argument that the singularity is physical.}
In Fig.~\ref{fig:chargeratio} we plot the ratio of the Maxwell charge at the horizon to the charge at infinity, and find that while it does decrease with increasing $\hat{\mu}$, it never crosses zero. Thus there does not appear to exist a finite-temperature negatively charged \textit{smeared} CGLP black brane, which is entirely consistent with expectations. 

Some novel features of the CGLP black brane relative to previous anti-brane investigations are the $AdS$ asymptotics and the existence of a dual field theory. It would be very interesting to understand the observed behaviour of the Maxwell charge from the CFT perspective, but unfortunately we do not understand the precise holographic dual of  $Q_{\text{Maxwell}}$. It is natural to associate the running of $Q_{\text{Maxwell}}$ with a series of duality cascades where the rank of the dual gauge group is decreased, but without a more complete understanding of the dual field theory this understanding remains only heuristic.  A connection of this nature between allowed charges of black branes in flux backgrounds and CFT quantities would be extremely useful.  One tantalising possibility is that if one better understood the role of the Maxwell charge in the dual field theory, then it might become obvious which types (if any) of negatively-charged black holes can exist in flux backgrounds with field theory duals.  We leave this interesting idea for future work, and will merely remark that in the dimensionally reduced theory, $Q_{\text{Maxwell}}$ is proportional to $\tilde{F}_0$, and so understanding the holographic dual of $\tilde{F}_0$ would be very useful. 

\section{Discussion \label{sec:discussion}}
\noindent\indent Asymptotically $AdS_4 \times V_{5,2}$ black branes in 11-dimensional supergravity provide a fertile ground for theoretical study, encompassing such diverse topics as novel black brane solutions in higher dimensions, thermodynamics of 2+1 dimensional CFTs, and the KKLT proposal for building de Sitter solutions in string theory. In this work we have tried to emphasise the connections between these different perspectives in the context of black brane solutions in the CGLP flux background.  Of the many black hole solutions that may exist, we have considered the simplest one: namely, the ``smeared'' black brane with the same internal $SO(5)$ symmetry as the zero-temperature CGLP solution.

By analogy with black holes in $AdS_5 \times S^5$ \cite{Dias:2015pda, Dias:2016eto}, one might expect a first-order confinement/de-confinement phase transition between the smeared CGLP black brane and the thermal CGLP solution.  Surprisingly, this does not happen.  Instead the family of CGLP black branes terminates in a singular solution, which appears to have precisely the same free energy as the confined CGLP solution---in other words, the expected phase transition would appear to coincide with the singularity!  There are a number of interesting ways this may be resolved.  Just as in $AdS_5 \times S^5$, we expect there to exist a variety of other solutions such as localised black branes, although these will be more difficult to construct \cite{Dias:2015pda, Dias:2016eto}.  One possible resolution to the puzzle is that these localised solutions participate in the confinement/de-confinement transition in the canonical ensemble in some way.  From the perspective of the field theory, these less-symmetric solutions correspond to states with broken $R$-symmetry, and it might be that confinement and spontaneous symmetry breaking are somehow related in this theory.  Another possibility is that one might fail to see a phase transition even after considering these less-symmetric solutions, and in the strict large-$N$ limit the theory does not confine. Presumably $1/N$ effects would then lead to confinement.

It is interesting to contemplate the different possible black hole phases that may exist. In the $AdS_5 \times S^5$ system, the expected localised solution with horizon topology $S^8$ was found, and in addition evidence was seen for an infinite family of solutions associated with different spherical harmonics on the $S^5$ (black branes for the $\ell = 1,2$ harmonics were explicitly constructed \cite{Dias:2016eto}). In the CGLP case, we certainly expect localised black branes to exist with horizon topology $\mathbb{R}^2 \times S^7$. There may or may not be an infinite number of solutions associated with harmonics on the $V_{5,2}$, it is difficult to say at this point. However, our geometric understanding of the singularity of the critical black brane suggests that a new family of solutions should connect to the family we have constructed here, with the two families joined by a topology changing transition at the singular point. The geometry of the singular solution suggests that the horizon topology of this new family may be $\mathbb{R}^2 \times S^3 \times S^4$. If such a solution exists, an interesting question is whether or not it plays a role in resolving the puzzle of the confinement/de-confinement transition.

Regarding the problem of anti-branes in flux backgrounds, we have demonstrated that smeared CGLP black branes cannot support negative charge on the horizon. This was entirely expected, and conforms to widely-held expectations as well as previous results for the CGLP background. Whether or not this is the case for more general black branes (including localised anti-branes) is an entirely different, and much more difficult problem with less consensus. While we do not have any concrete results for this problem as of yet, we end with a few concluding remarks on the subject. In Ref.~\cite{Hartnett:2015oda}, it was argued that solutions with less spatial symmetry have greater freedom to support negative charge at the horizon. In the CGLP background, any less symmetric solutions such as a black brane with horizon topology $\mathbb{R}^2 \times S^3 \times S^4$ or $\mathbb{R}^2 \times S^7$ may well be able to support negative charge at the horizon. Additionally, if the supergravity equations of motion impose some restriction of the charge of black branes in the CGLP background, then there should be a corresponding statement in the dual field theory. This is a very intriguing possibility. By examining the details of the smeared black brane constructed here, it seems that the relevant quantity on the gravity side is the Maxwell charge, or equivalently the 4-dimensional 0-form flux $\tilde{F}_0$. An understanding of the holographic interpretation of either of these quantities could shed light on what restrictions anti-branes in flux are subject to.

\begin{acknowledgments}
We wish to thank Iosif Bena, Kostas Skenderis, and Fri{\dh}rik Gautason for useful discussions.   The work of O.D. and G.S.H. is supported by the STFC Ernest Rutherford grants ST/K005391/1 and ST/M004147/1. The work of BEN has been supported variously by ERC grant ERC-2011-StG 279363-HiDGR and by ERC grant ERC-2013-CoG 616732-HoloQosmos.   We gratefully acknowledge the computer resources, technical expertise and assistance provided by CENTRA/IST. Some computations were performed at the cluster ``Baltasar-Sete-S\'ois'' and supported by the H2020 ERC Consolidator Grant ``Matter and strong field gravity: New frontiers in Einstein's theory'' grant agreement no. MaGRaTh-646597.
\end{acknowledgments}

\appendix
\section{The DeTurck trick in a non-coordinate basis}
\label{app:noncoord}
\noindent\indent In Sec.~\ref{subsec:deturck} we have introduced the DeTurck trick for rendering the Einstein equation elliptic (for time-independent solutions) by making its gauge degrees of freedom dynamical, in such a way that solutions of the Einstein-DeTurck equation correspond in the end to solutions of the original Einstein equation in the DeTurck gauge \cite{Headrick:2009pv}.  This trick involves adding the Lie derivative $\Lie_\xi g$ to the Einstein equation for a particular vector field $\xi$ which is typically written in either of the following ways:
\begin{equation} \label{eq:xi-coord-def}
\xi^\mu = g^{\nu \rho} \Big( \Gamma^\mu_{\nu \rho} - \bar \Gamma^\mu_{\nu \rho} \Big), \quad \text{or} \quad \xi_\mu = g^{\nu \rho} \Big( \bar{\nabla}_\nu g_{\mu \rho} - \frac12 \bar \nabla_\mu g_{\nu \rho} \Big),
\end{equation}
where the barred quantities refer to a given \emph{reference metric}.  Conventions in general relativity typically hold that Greek indices from the middle of the alphabet refer to a coordinate basis $\partial / \partial x^\mu$, and when explicit Christoffel symbols appear, one may have concerns as to whether the expression written is indeed applicable in a non-coordinate basis or not.  Since in this work we have used a collection of left-invariant forms rather than a coordinate basis, it is useful to clearly develop the DeTurck method for such bases.

The starting point is the \emph{second} expression in \eqref{eq:xi-coord-def}, in which each individual term is obviously covariant.  This expression is simply a certain combination of traces of the \emph{non-metricity tensor} $N^{\bar \nabla}$ for the reference connection $\bar \nabla$.  The non-metricity tensor of a given connection $\bar \nabla$ is just the covariant derivative of the metric,
\begin{equation} \label{eq:non-metricity-coord}
N^{\bar \nabla}_{\mu \nu \rho} \equiv \bar \nabla_\mu g_{\nu \rho},
\end{equation}
and of course for a metric-compatible connection it vanishes.  But since this is a tensorial object, we can readily change it into any basis, including non-coordinate ones.  A more basis-agnostic way of writing the non-metricity tensor $N^{\bar \nabla} : TM \times TM \times TM \to \RR$ is in terms of its action on three arbitrary vector fields $X, Y, Z$:
\begin{equation}
N^{\bar \nabla}_Z (X,Y) = \bar \nabla_Z g(X, Y) - g( \bar \nabla_Z X, Y) - g(X, \bar \nabla_Z Y).
\end{equation}
Of course one easily recovers the components in a basis by choosing $X, Y, Z$ from, e.g., some convenient set of frame fields $e_A$.  Using the metric tensor $g \equiv g_{AB} \, e^A \otimes e^B$, one can then define the vector $\xi$ via its inner product with an arbitrary vector $X$:
\begin{equation}
g(\xi, X) = g^{AB} \Big[ N^{\bar \nabla}_{e_A} (e_B, X) - \frac12 N^{\bar \nabla}_X (e_A, e_B) \Big],
\end{equation}
or equivalently, ``raising an index'', we can obtain the components of $\xi^A$ itself:
\begin{equation} \label{vec-deturck}
\xi^A = g^{AD} g^{BC} \Big[ N^{\bar \nabla}_{e_B} (e_C, e_D) - \frac12 N^{\bar \nabla}_{e_D} (e_B, e_C) \Big].
\end{equation}

In order to \emph{obtain} the non-metricity tensor for the types of calculations we do in this work, it is convenient to choose an \emph{orthogonal} frame rather than an orthonormal one.  The reason for this is that the reference metric $\bar g$ uses the \emph{same} 1-forms $\sigma_i, \tilde \sigma_i, \nu$ to represent the isometry directions, but puts different functions in front of them; e.g. on the Stiefel part of the metric ansatz one has
\begin{equation}
g = a(r)^2 \, \sigma_i^2 + b(r)^2 \, \tilde \sigma_i^2 + c(r)^2 \, \nu^2, \qquad \bar g = \bar a(r)^2 \, \sigma_i^2 + \bar b(r)^2 \, \tilde \sigma_i^2 + \bar c(r)^2 \, \nu^2,
\end{equation}
where in the reference metric, the barred functions are fixed \emph{a priori}.  Since we wish to obtain $N^{\bar \nabla} \equiv \bar \nabla g$, it is useful to use the orthogonal basis
\begin{equation} \label{orthogonal-basis}
e^\nu = \nu, \qquad e^i = \sigma_i, \qquad e^{\tilde \imath} = \tilde \sigma_i.
\end{equation}
Then one has $\bar \nabla e^A \equiv - \bar \omega^A{}_B \otimes e^B$ for the connection 1-forms $\bar \omega^A{}_B$.  Since $\bar \nabla$ is Levi-Civit\`a \emph{with respect to $\bar g$}, these connection 1-forms can be obtained by solving the metric-compatibility (for the \emph{reference} metric $\bar g$) and torsion-free conditions given by:
\begin{align}
\bar \nabla \bar g &= 0 = \Big( \dd \bar g_{A B} - \bar g_{A C} \, \bar \omega^{C}{}_{B} - \bar g_{B C} \, \bar \omega^{C}{}_{A} \Big) \otimes e^{A} \otimes e^{B}, \label{metric-compat} \\
\bar T^A &= 0 = \dd e^A + \bar \omega^A{}_B \wedge e^B.
\end{align}
Note that since the frame is \emph{not} orthonormal, the metric coefficients $\bar g_{AB} \neq \delta_{AB}$, and \eqref{metric-compat} replaces the usual antisymmetry condition one would have in an orthonormal frame.  Having obtained the connection 1-forms $\bar \omega^A{}_B$, one can then evaluate $\bar \nabla g$ and take the traces in \eqref{vec-deturck} to obtain the DeTurck vector $\xi$.

Finally, it is instructive to re-write the covariant expression \eqref{vec-deturck} in terms of the connection 1-forms $\omega^A{}_B$ and $\bar \omega^A{}_B$ (of $\nabla$ and $\bar \nabla$, respectively) in the (common) orthogonal basis \eqref{orthogonal-basis}.  This exercise in index gymnastics yields
\begin{equation} \label{eq:xi-noncoord-def}
\xi^A = g^{BC} \Big( \omega_B{}^A{}_C - \bar \omega_B{}^A{}_C \Big) + g^{AC} \Big( \omega_B{}^B{}_C - \bar \omega_B{}^B{}_C \Big) - g^{AC} \Big( \omega_C{}^B{}_B - \bar \omega_C{}^B{}_B \Big).
\end{equation}
Comparing this to the \emph{first} expression in \eqref{eq:xi-coord-def}, we can see that we were right to view it with suspicion.  The change to the non-coordinate basis has not merely introduced a substitution $\Gamma \to \omega$, but has generated additional terms.  In fact, we ought to have expected this, since $\Gamma^\mu_{\nu \rho}$ is symmetric in its lower indices, while $\omega_B{}^A{}_C$ is not, and thus the simple operation of tracing over the lower indices is removing information from $\omega_B{}^A{}_C$ that was not removed in the expression of \eqref{eq:xi-coord-def}.

The expression obtained in \eqref{eq:xi-noncoord-def} is in fact valid in \emph{any} basis, be it orthogonal or not, and thus should be viewed as the generalization of the first expression in \eqref{vec-deturck} for generic bases.  A word of caution, however:  One must ensure that the $A, B$ indices of $\omega^A{}_B$ and $\bar \omega^A{}_B$ refer to the \emph{same} basis; that is, one should not mistakenly think of $\omega^A{}_B$ and $\bar \omega^A{}_B$ as the connection 1-forms for orthonormal bases on $g$ and $\bar g$.  Alternatively, one can stick to the manifestly covariant expression \eqref{vec-deturck}.


\bibliography{masterrefs}
\bibliographystyle{utphys}

\end{document}